\documentstyle[preprint,eqsecnum,prd,aps]{revtex}
\begin{document} 
% states and fields

%%%%Start of Text%%%%%%%%%%%%%%%%%%%%%%%%%%%%%%%%%%%%%%%%%%%%%%%%%%%%%%%%%%%%
%\rightline{
\preprint{
\vbox{
\halign{&##\hfil\cr
	& ANL-HEP-PR-95-87 \cr
	& EC-HEP-960201 \cr
	& February 15, 1996 \cr}}
}
%}
\title{Production of a Prompt Photon in Association with Charm 
at Next-to-Leading Order in QCD}

\author{B. Bailey$^a$, Edmond L. Berger$^b$, and L. E. Gordon$^b$}
\address{$^a$ Physics Department, Eckerd College, St. Petersburg, FL
33711\\
$^b$High Energy Physics Division, Argonne National Laboratory,
	Argonne, IL 60439}
\maketitle
\begin{abstract} 
A second order, $O(\alpha ^2_s)$, calculation in perturbative quantum 
chromodynamics of the two particle inclusive cross section is presented 
for the reaction $p +\bar{p}\rightarrow \gamma + c + X$
for large values of the transverse momentum of the prompt photon and charm 
quark.  The combination of 
analytic and Monte Carlo integration methods used here to perform phase-space
integrations facilitates imposition of photon isolation restrictions and other
selections of relevance in experiments.  Differential distributions are 
provided for various observables.  Positive correlations in rapidity are 
predicted.  
%\end{abstract}
\vspace{0.2in}
\end{abstract}
\narrowtext

\section{Introduction}
More precise examination of the expectations of perturbative quantum
chromodynamics(QCD), including dynamical correlations inherent in the 
hard-scattering matrix elements, is made possible by the observation of 
inclusive production of two particles or jets each carrying a large value of 
transverse momentum.  Because they couple in a point-like fashion to quarks,
the observation of photons with large values of transverse momentum in a high 
energy hadron collision has long been regarded as an incisive probe of short
distance dynamics.  In the case of inclusive production of heavy quarks, the
large mass of the quark and/or the fact that the quark carries large transverse 
momentum justifies use of a perturbative short-distance approach.  In this
paper, we continue our examination of the associated production of a prompt
photon along with a heavy quark\cite{bergergordon}.  Data are beginning to 
become available on the associated production of a photon $\gamma$  
carrying large transverse momentum along with a charm quark $c$ whose 
transverse momentum balances a substantial portion of that of the 
photon\cite{cdf}.  An intriguing possibility is that the data may be used to 
measure the charm quark density in the nucleon.

In this paper we report results of a full next-to-leading order 
perturbative QCD calculation of $p +\bar{p}\rightarrow \gamma + c + X$
at high energy.  For values of the transverse momentum $p_T^c$ of the 
charm quark much larger than the mass $m_c$ of the quark, only one 
{\it {direct}} hard scattering subprocess contributes in leading order: 
the quark-gluon Compton subprocess $g c \rightarrow \gamma c$.  The initial 
charm quark and the initial gluon are constituents of the initial hadrons.  In 
addition, there is a leading order {\it {fragmentation}} process in which the
photon is produced from quark or gluon fragmentation, e.g.,
$g g \rightarrow c \bar{c}$ followed by $\bar{c} \rightarrow \gamma X$, or
$q c \rightarrow  q c$ followed by $q \rightarrow \gamma$.
At next-to-leading order in QCD, several subprocesses contribute to the
$\gamma + c +X$ final state: $gc \rightarrow gc\gamma$,
$g g \rightarrow c \bar{c} \gamma$,
$q \bar{q} \rightarrow c \bar{c} \gamma$,
$q c \rightarrow  q c \gamma$,
$\bar{q} c \rightarrow \bar{q} c \gamma$,
$c \bar{c} \rightarrow c \bar{c} \gamma$, and
$c c \rightarrow c c \gamma$.
A full next-to-leading order calculation requires the computation of the 
hard-scattering matrix elements for these two-to-three particle production 
processes as well as the one-loop $O(\alpha_s^2)$ corrections to the lowest 
order subprocess $g c \rightarrow \gamma c$.   

We are interested ultimately in the fully differential two-particle
inclusive cross section,  $E_\gamma E_cd\sigma/d^3p_\gamma d^3p_c$, where
$(E,p)$ represents the four-vector momentum of the $\gamma$ or $c$ quark.  For 
each contributing subprocess, this calculation requires integration over the
momentum of the unobserved final parton in the two-to-three particle 
subprocesses
($g$, $\bar{c},q$, or $\bar{q}$).  Collinear singularities must be handled
analytically by dimensional regularization and absorbed into parton momentum
densities or fragmentation functions.  In the theoretical analysis reported 
here, a combination of analytic and Monte Carlo integration methods is used to 
perform phase-space integrations over unobserved final-state partons and the
momenta of the initial partons.  This approach facilitates imposition of 
photon isolation restrictions and other selections of relevance in experiments. 
We work in the massless approximation, $m_c = 0$.  To warrant use of 
perturbation theory and the massless approximation, we
limit our considerations to values of transverse momenta of the photon
and charm quark $p^{\gamma,c}_{T} > 10$ GeV.  

In the lowest order direct subprocess, $g c \rightarrow \gamma c$, the prompt 
photon emerges in isolation from the only other particle in the hard 
scattering, the charm quark.  Long-distance quark-to-photon
and gluon-to-photon fragmentation processes have been
emphasized theoretically\cite{field} and parametrized phenomenologically in
leading order\cite{frag1}, and evolved in next-to-leading
order\cite{frag2,frag3}. These terms may account for more than half of the
calculated inclusive single photon cross section at modest values of transverse
momentum at the Fermilab Tevatron collider.  Photons originating through
fragmentation are likely to emerge in the neighborhood of associated hadrons.
An experimental
isolation restriction is needed before a clean identification can be made of
the photon and a measurement made of its momentum.  Isolation reduces the size
of the observed fragmentation contribution.  Photon
isolation complicates the theoretical interpretation of results, however, since
it threatens to upset the cancellation of infra-red divergences in perturbation
theory\cite{berqiu}. In this paper, we include the fragmentation contributions, 
and we impose isolation requirements through our Monte Carlo 
method.  

A combination of analytic and Monte Carlo methods similar to that we employ 
in this paper has been used to carry out next-to-leading order QCD 
calculations of other processes including inclusive 
prompt photon\cite{baer1} and photon pair \cite{bailey1} production in hadron 
collisions, single \cite{baer2} and pair production of heavy gauge 
bosons \cite{ohnemus}, and heavy flavor pair production \cite{correl}.

In Section II, we describe the combination of analytic and Monte Carlo methods
we use to carry out the next-to-leading order calculation.  The 
next-to-leading order calculation itself is presented in Section III.  
Differential cross sections and other numerical results  
are discussed in Section IV.  Summary remarks are collected in
Section V.  An appendix is included in which we derive analytic expressions 
for some of the parton level cross sections we use.  

\section{Monte Carlo Method}

The combination of analytic and Monte Carlo techniques used here to
perform the phase space integrals is documented and described in
detail elsewhere \cite{baer1,bailey1,baer2,ohnemus}, so 
our discussion will be fairly brief, highlighting features important to our
calculation. For the two-to-three particle hard-scattering subprocesses, the 
technique consists in identifying those
regions of phase space where soft and collinear singularities occur and
integrating over them analytically in $4-2\epsilon$ dimensions. In this
way the singularities are exposed as poles in $\epsilon$. These regions
are isolated from the rest of the three-body phase space by the imposition
of arbitrary boundaries through the introduction of cutoff
parameters, $\delta_s$ and $\delta_c$. The soft gluon region of phase space is 
defined to be the
region in which the gluon energy, in a specified reference frame, usually
the subprocess rest frame, falls below a certain threshold,
$\delta_s\sqrt{\hat{s}}/2$, where $\delta_s$ is the cutoff
parameter, and $\hat{s}$ is the center-of-mass energy in the initial
parton-parton system. Labelling the momenta for the general three-body
process by $p_1+p_2\rightarrow p_3+p_4+p_5$, we define the
general invariants by $s_{ij}=(p_i+p_j)^2$ and $t_{ij}=(p_i-p_j)^2$. The
collinear region is defined as the region in which the value of an invariant
falls below the value $\delta_c \hat{s}$. 

The full set of three-body final-state subprocesses is:
\begin{mathletters}\label{eq:1}
\begin{eqnarray}
g &+ c \rightarrow g + c + \gamma\label{eq:11}\\
g &+ g \rightarrow c +\bar{c} + \gamma\label{eq:12}\\
q &+ \bar{q} \rightarrow c +\bar{c} + \gamma\label{eq:13}\\
q &+ c \rightarrow  q + c + \gamma\label{eq:14}\\
\bar{q} &+ c \rightarrow \bar{q} + c + \gamma\label{eq:15}\\
c &+ \bar{c} \rightarrow c + \bar{c} + \gamma\label{eq:16}\\
c &+ c \rightarrow c + c + \gamma\label{eq:17}
\end{eqnarray}
\end{mathletters} 

The matrix elements integrated over the mutually exclusive soft and
collinear regions of phase space are not the full two-to-three body matrix 
elements but, instead, 
specific approximate versions. In the soft-gluon case the approximate version 
is obtained by setting the gluon energy to zero everywhere 
it occurs in the matrix elements, except in the denominators.  For the 
collinear singularities, each 
invariant that vanishes is in turn set to zero everywhere except in the
denominator. This form is the leading pole approximation.  The phase space
integrals are performed with these approximate expressions, and only the terms 
proportional to logarithms of the cutoff parameters are retained.  Terms 
proportional to positive powers of the cutoff
parameters are set to zero.  In order for the method to yield reliable 
results, the cutoff parameters must be kept small, otherwise the 
approximations would not be valid.  

After the two-to-three particle phase space integrals are performed 
analytically over the singular regions and the soft and collinear poles are 
exposed, the $O(\alpha ^2_s)$ virtual gluon-exchange loop contributions, if 
any, are added, and all double poles and single poles of soft (infrared)
origin are verified to cancel, as they should. The remaining collinear 
singularities are factored into parton distribution and fragmentation 
functions at an appropriate factorization or fragmentation scale.
We work in the $\overline{MS}$ scheme. One is left with a set of matrix 
elements for effective two-body final-state processes that depend explicitly 
on $\ln\delta_s$ and $\ln \delta_c$.  In addition, the non-singular regions of 
phase space yield a set of three-body final-state matrix elements
which, when integrated over phase space by Monte Carlo techniques,
have an implicit dependence on these same logarithms.  The signs are such 
that the dependences on $\ln\delta_s$ and $\ln \delta_c$ cancel between 
the two-body and three-body contributions. The physical cross sections are
independent of these arbitrary cutoff parameters over wide ranges.  In our 
numerical work, we 
varied $\delta_s$ and $\delta_c$ over suitable ranges and found quite stable
results, as is shown in Sec. IV.  

At the level of two-body final-state matrix elements, as in leading-order 
calculations, it is a simple matter to impose selections on kinematic 
variables similar to those made in experiments and to calculate
different observables.  The same is not the case when we consider three-body 
final-state processes.  The standard analytic techniques required to obtain
differential cross sections of empirical interest often involve
complex Jacobian transformations, and the phase space integrals can 
sometimes be done analytically only when specific limits of integration
are involved.  Fully analytic methods of performing
calculations for physical processes, although in some cases desirable,
can be rather restricted in their usefulness when it is desirable and sometimes
even unavoidable that kinematic selections be made to model experimental 
cuts. The combined analytic and Monte Carlo method is particularly versatile 
in that it provides a means to calculate cross sections differential in many 
variables at once, and to apply cuts on the kinematic variables to match those
made in the experiments.   The phase space 
integrals are performed numerically after all singularities have been 
handled analytically.  

Our calculation of photon plus charm quark production proceeds along lines 
similar to that for inclusive direct photon
production described in Ref.~\cite{baer1}, but with a few important differences.
Since we are interested in observing a final charm quark as well as the photon,
we cannot integrate analytically over phase space in the limit in
which the charm quark is collinear to a hard gluon or an anti-charm quark. This
situation occurs in the $cg$, $c\bar{c}$, and $q\bar{q}$ initiated
processes of Eq.~(\ref{eq:1}). The expression in the appendix of 
Ref.~\cite{baer1} 
for the sum of all effective two-body contributions cannot be used in our 
calculation.  We recalculated this expression using the three-body matrix 
elements and virtual gluon-exchange contributions from Ref.~\cite{gorvogel}, 
and we provide the results in our appendix. We also discuss and present in the
appendix the final-state collinear remnants that give the charm quark 
momentum distribution in the limit that the charm quark is produced 
collinearly with an anti-charm quark or a gluon. The singularities in these 
cases are factored into charm quark fragmentation functions.          

\section{Contributions Through Next-to-Leading Order }

\subsection{Leading order contributions}

In leading order in perturbative QCD, only one {\it{direct}} subprocess
contributes to the hard-scattering cross section, the QCD Compton process
$c g\rightarrow \gamma c$,
unlike the case for single inclusive prompt photon production, where the
annihilation process $q\bar{q}\rightarrow \gamma g$ also contributes.
Since the leading order direct partonic subprocess has a two-body final state,
the photon and $c$ quark are produced with balancing transverse
momenta.  In addition, there are effectively leading-order contributions in 
which the photon is produced by fragmentation from a final-state parton. These 
are
\begin{eqnarray}
c+g&\rightarrow& g+c \nonumber \\
g+g&\rightarrow& c+\bar{c} \nonumber \\
c+q&\rightarrow& c+q \nonumber \\
c+\bar{q}&\rightarrow&c+\bar{q} \nonumber \\
c+c&\rightarrow& c+c \nonumber \\
c+\bar{c}&\rightarrow & c+\bar{c} \nonumber \\
q+\bar{q}&\rightarrow & c+\bar{c}.
\label{eq:fragproc}
\end{eqnarray}
If the photon is to be isolated from the observed charm quark, it arises from 
fragmentation of the gluon $g$ and the non-charm quark $q$, respectively,
in the cases of the first, third and fourth processes.  In the
other cases it is produced by fragmentation of one of the (anti)charm quarks.

In a fully consistent next-to-leading calculation, one should calculate
the subprocesses in Eq.~(\ref{eq:fragproc}) to $O(\alpha_s^3)$, since the photon
fragmentation functions that are convoluted with the hard subprocess
cross sections are of $O(\alpha_{em}/\alpha_s)$.  For simplicity, we
include them in $O(\alpha_s^2)$ only.  In fact, next-to-leading order
fragmentation contributions to single prompt photon production have been
included only once before\cite{gorvogel}. We expect the next-to-leading order
corrections to the fragmentation contributions to be insignificant numerically
especially after isolation cuts are imposed.

\subsection{Next-to-leading order contributions}

There are two classes of contributions in next-to-leading order. First there
are the virtual gluon exchange corrections to the lowest order process, 
$c g\rightarrow \gamma c$. Examples are
shown in Fig.1(b). These amplitudes interfere with the Born amplitudes and
contribute at $O(\alpha_{em}\alpha_s^2)$. They were calculated
twice before\cite{baer1,gorvogel}.  At next-to-leading order there are also
three-body final-state contributions, listed in Eq.~(\ref{eq:1}). The
matrix elements for these are also taken from Ref.~\cite{gorvogel}, where 
they are calculated for single inclusive prompt photon production.

The main task of our calculation is to integrate the three-body matrix elements
over the phase space of
the unobserved particle in the final state. The situation here is
different from the standard case of single inclusive particle
production because we wish to
retain as much control as possible over the kinematic variables of a
second particle in the final state, while at the same time integrating
over enough of the phase space to ensure cancellation of all infrared
and collinear divergences, inherent when massless
particles are assumed.  All the processes of Eq.~(\ref{eq:1}), except the 
first, involve collinear singularities but no soft singularities. These 
collinear 
singularities must be exposed and factored as explained in Sec.~II. The 
results of these calculations are listed in the appendix. 

At $O(\alpha ^2_s)$ there are, in addition, fragmentation processes
in which the hard-scattering two-particle final-state subprocesses
\begin{eqnarray}
c+g &\rightarrow& \gamma+ c \nonumber \\
c+\bar{c}&\rightarrow& \gamma+g \nonumber \\
q+\bar{q}&\rightarrow& \gamma +g
\end{eqnarray}
are followed by fragmentation processes $c\rightarrow c X$, in the case of
the first subprocess, and
$g\rightarrow c X$ in the cases of the last two. These should be included
because we have factored the collinear singularities in the
corresponding three-body final-state processes into non-perturbative 
fragmentation functions for production of a charm quark from a particular 
parton. As a first approximation, we estimate these fragmentation functions by
\begin{eqnarray}
D_{c/c}(z,\mu^2)&=&\frac{\alpha_s(\mu^2)}{2\pi}P_{qq}(z),
\label{eq:cfrag}
\end{eqnarray}
and
\begin{eqnarray}
D_{c/g}(z,\mu^2)&=&\frac{\alpha_s(\mu^2)}{2\pi}P_{qg}(z),
\label{eq:gfrag}
\end{eqnarray}
where $P_{ij}(z)$ are the lowest order splitting functions for parton
$j$ into parton $i$ \cite{altar}; and $\alpha_s(\mu^2)$ is the strong
coupling strength. 

\section{Numerical Results}

In this section we present and discuss several differential
cross sections for the joint production of a charm quark and a 
photon at large values of transverse momentum. All results are displayed
for $p\bar{p}$ collisions at the center-of-mass
energy $\sqrt{s}=1.8$ TeV appropriate for the CDF and D0 experiments
at Fermilab.  To obtain the differential cross sections presented in this 
paper, we convolute our hard-scattering matrix elements with the CTEQ3M parton 
densities\cite{parden2}.  Very similar differential distributions may be 
obtained if other parton sets are used instead, with quantitative differences 
reflecting differences among charm quark densities in the different 
sets\cite{bergergordon}.  We set the renormalization, factorization, and 
fragmentation scales to a common value $\mu = p_T^{\gamma}$ in most
of our calculations.  Dependence on $\mu$ is examined in one of the 
figures below.  Since there are two particles in the final state,
the charm quark and the photon, both of whose transverse momenta are
large, an alternative choice might be $\mu = p_T^c$ or some function
of $p_T^{\gamma}$ and $p_T^c$.  The results of our calculations
show that the magnitudes of $p_T^{\gamma}$ and $p_T^c$ tend to be 
comparable and that dependence of the cross sections on $\mu$ is slight.  
Therefore, choices of $\mu$ different from $\mu = p_T^{\gamma}$ should not 
produce significantly different answers, and we have verified this 
supposition in representative cases.

In addition to showing distributions in the rapidities and transverse
momenta of the charm quark and the photon, we also discuss distributions in 
the ratio $z$, where z is defined as
\begin{equation}
z = - {{p^c_{T}. p^{\gamma}_{T}}\over {(p^\gamma_{T})^2}}.  \label{eq:zdef}
\end{equation}
This ratio $z$ is not to be confused with the variable $z$ in the fragmentation
functions, Eqs.~(\ref{eq:cfrag}) and (\ref{eq:gfrag}).   
In collider experiments a photon is observed and its momentum is well
measured only when the photon is isolated from neighboring hadrons.  In
our calculation, we impose isolation in terms of the cone variable R:  
\begin{equation}
\sqrt{(\Delta y)^2 + (\Delta \phi)^2} \leq R.     \label{eq:Rdef}
\end{equation}
In Eq.~(\ref{eq:Rdef}), $\Delta y$ ($\Delta \phi$) is the difference between 
the rapidity (azimuthal angle in the transverse plane) 
of the photon and that of any parton in the final state.  The photon is said
to be isolated in a cone of size $R$ if the ratio of the hadronic energy in the 
cone and the transverse momentum of the photon does not exceed 
$\epsilon = 2 GeV/p_T^{\gamma}$. We show distributions
for the choices $R = 0.7$ and $R = 0.4$ typical of current experiments.  

Our first figure in this section, Fig. 2, is an examination of the numerical 
stability of the
overall cross section when the cutoff parameters $\delta_s$ and $\delta_c$
are varied over appropriate ranges.  This figure indicates that the combination
of analytic and Monte Carlo methods yields consistent 
numerical results for a broad range of the parameters.  For the subsequent
figures, we use $\delta_s =$0.01, and $\delta_c =$0.001.  We return to a brief 
examination of dependence on cutoff parameters when we discuss distributions in
the variable $z$ in Fig. 9.
 
In Fig. 3 we show the differential cross section as a function of the
transverse momentum of the charm quark $p_T^c$, having restricted the
transverse momentum of the photon to the range $15\leq
p^{\gamma}_T\leq 45$ GeV typical of current hadron collider experiments. The
rapidities of the charm quark and photon are restricted to the central
region, $-1\leq y^{\gamma,c}\leq 1$ in order to mimic the central region 
coverage of major collider detectors.  The solid curve
shows our prediction when no further selections are made other than those
mentioned just above.  Distributions are presented in the
figure for various selections on other kinematic variables.  For the dashed
curve, the variable $z$ of Eq.~(\ref{eq:zdef}) is restricted to 
$z\geq 0.1$.  This selection on $z$ places the photon and charm quark in
opposite hemispheres and results in a modest reduction in overall rate.
Retaining this cut on $z$, we examine the effects of isolation of the photon
and obtain the results shown by the dotted and dot-dashed curves, for cone
sizes of $R=0.4$ and $0.7$ respectively.  
 
A common and notable feature of the curves in Fig.3 is that slopes 
change near $p_T^c=15$ and $45$ GeV. There is a simple reason for this 
behavior.  The contributions to the cross section from two-particle final
states produce kinematic configurations in which the photon and charm quark 
have values of $p_T$ that are equal in
magnitude but opposite in sign.  Therefore two-body final-state processes 
cannot contribute in the regions $p_T^c\leq 15$ GeV and $p^c_T\geq 45$ GeV.
Only the three-body final-state processes contribute to the cross section
in these regions.  The steeper fall of the cross section in either direction 
away from the region $15\leq p^c_T\leq 45$ GeV reflects the decreasing
likelihood that the photon and charm quark have substantially different values
of transverse momentum.  

Another feature of the results shown in Fig. 3 is that the effect of isolation 
diminishes as $p^c_T$ is decreased.  Isolation affects the cross section 
principally when the third parton in final state enters the photon 
isolation cone and carries transverse momentum greater than the energy 
resolution threshold.  With $p^\gamma_T$ fixed above $15$ GeV,
the third parton must be in the charm quark's hemisphere when $p_T^c$ is small
in order to balance $p^{\gamma}_T$.  When $p_T^c$ is large, 
the third parton is free to enter the photon isolation cone. 

In Fig. 4, we show the cross section differential in $p_T^\gamma$. 
The cuts made are the same as those for Fig. 3, but in this case the
charm quark's transverse momentum is restricted between $15$ and $45$ GeV. 
The explanation for the change in behavior of the distributions 
above $p_T^\gamma = 45$ GeV and below $p_T^\gamma =15$ GeV is the same as for 
Fig. 3.  An obvious difference between Figs. 3 and 4 is that 
the effect of photon isolation is most significant in the region of small
$p^\gamma_T$.  The explanation is, again, that with $p_T^c$
restricted above $15$ GeV, the third parton in the final state will be
found in the photon hemisphere when $p_T^\gamma$ is small. It is likely 
to be in the photon
isolation cone, and the configuration will be rejected by the isolation cuts. 

To examine further the effects of selections on the charm quark momentum,
we present in Fig. 5 the cross section differential in the transverse 
momentum of the photon for a different set of cuts.  The photon's rapidity 
is limited to the range $-0.5\leq y^\gamma \leq 0.5$, and the ratio $z$ is
restricted to $0.2\leq z \leq 2.0$.  These cuts are similar to those of
our analytic paper\cite{bergergordon}.  In Fig. 5, the
solid curve represents the cross section with no isolation cuts imposed,
and the dashed curve shows the isolated cross section. The dot-dashed
curve is the leading order prediction, with photon isolation imposed.
The behavior seen in Fig. 5 is clearly different from that of Fig. 4 in
that the cross section does not fall off in the region of small $p^\gamma_T$,
as expected, since there in no selection in Fig. 5 on $p^c_T$ (other than the 
selection on $z$).

In Fig. 6 we show the distribution in the rapidity of the charm quark, $y^c$,
for different cuts on the photon's transverse momentum. In all cases $-1\leq
y^\gamma\leq 1$.  The distribution in $y^c$ is fairly broad, with full-width
at half-maximum of about 3.2 units in rapidity.  The dashed and 
dot-dashed curves show that the distribution in $y^c$ may broaden 
somewhat as $p_T^{\gamma}$ is increased.

The structure of the QCD hard-scattering matrix element produces 
{\it {positive}} correlations in rapidity\cite{elbcor} at collider energies. 
To examine correlations more precisely, we study 
the cross section as a function of the difference of the rapidities of the 
photon and charm quark.  Results are shown in Fig. 7.  In Fig. 7(a), we see
that the distribution in $\Delta y$ is narrower than the corresponding 
distribution in $y^c$ shown in Fig. 6.  The broader distribution in Fig. 6
results from a spread of the approximately Gaussian and relatively narrow 
dynamical distribution of Fig. 7(a) over the range $-1\leq y^\gamma\leq 1$.  
In Fig. 7(b), we select photons whose rapidities are in the forward
hemisphere, $1.0 < y^{\gamma} < 2.0$.  We observe that the peak in the 
$\Delta y$ distribution remains close to $\Delta y = 0$, reflecting
the predicted\cite{elbcor} positive dynamical correlations, but with the 
typical 
value of $y^c$ lagging somewhat behind that of the selected $y^{\gamma}$.   

In Fig. 8, we display the differential cross section in $y^c$ itself, for 
two intervals of $y^{\gamma}$ in the forward rapidity region.  These
distributions show how the typical rapidity of the charm quark follows that
of the photon.   

The dependence of the cross section on the variable $z$, defined in 
Eq.~(\ref{eq:zdef}), is indicative of the imbalance in transverse momentum of
the charm quark and the photon.  For two-body processes, such as the 
leading-order Compton subprocess $g c \rightarrow \gamma c$, 
the photon and charm quark have balancing transverse momenta, and the
distribution is a $\delta$-function in $z$, $\delta(1-z)$. 
Contributions away from $z=1$ are due to the higher
order three-body contributions or to fragmentation processes. As discussed
in Ref.~\cite{bergergordon}, the photon fragmentation processes contribute in 
the region $z\geq 1$ only.  Processes in which the charm quark is
produced via fragmentation contribute in the region $0\leq z\leq 1$.
We thus expect that the effect of photon isolation will be observed in 
the region $z\geq 1$.   This expectation is confirmed in the results 
of Fig. 9(a) that show the cross section as a function of $z$ for 
the non-isolated and isolated cases.  The solid histogram in Fig. 9(a) agrees
quantitatively with the corresponding histogram in our analytic 
paper\cite{bergergordon} except for differences associated with the different
choice of parton densities.  

The $\delta$-function behavior at leading order is, of course, moderated 
by non-perturbative effects associated with ``intrinsic" transverse momentum
of the initial partons as well as by next-to-leading order perturbative 
contributions.
In this paper, we are working in the usual purely perturbative framework in 
which the initial partons are assumed to be collinear.  For a three-parton final
state, the region of $z$ near unity is the region in which one of the three
final partons becomes soft.  
Sensitivity to soft-gluon effects and the necessity for resummation procedures
is a common limitation when one considers next-to-leading order contributions
to an observable that is proportional to a $\delta$-function in 
leading order.  In our
calculation the soft gluon corrections to the three-body processes are
considered as effective two-body contributions, as discussed in section II. 
These contribute to the cross section at $z=1$, meaning that
all dependence on the soft cutoff parameter $\delta_s$ is concentrated at $z=1$.
In Fig. 9(b) we show the distribution in $z$ for 
different values of $\delta_s$.  The cross section is fairly independent of 
this parameter except when it becomes larger than about $0.02$. Similarly, in 
Fig. 9(c), we examine dependence on $\delta_c$, the collinear cutoff parameter.
We find fair stability over a reasonable range of values of $\delta_c$.
The observed $\delta_c$ variation is similar to the scale dependence seen in
Fig.~6 of Ref.~\cite{bergergordon}. The two have a related physical origin.
In the approach of Ref.~\cite{bergergordon}, the choice of a scale
$\mu$ partitions the calculation arbitrarily into
two-body collinear and three-body final-state contributions.  If
$\mu$ is increased, more of the next-to-leading order QCD contributions are
placed into the parton distributions (collinear kinematics),
and less into the exact three-body kinematics. Likewise, here the parameter
$\delta_c$ partitions the overall QCD matrix element into two-body collinear
and three-body final-state contributions.  It therefore is not a surprise to
see the same type of variation with $\delta_c$ in Fig.~9(c) as is exhibited
by scale variation.

The infrared sensitivity of the distribution in $z$, reflected in the $\delta_s$
dependence discussed above, and in the scale dependence examined in 
Ref.~\cite{bergergordon}
means that the $z$ distribution at may not be calculated sufficiently 
reliably at next-to-leading order.  Resummation of the effects of soft gluon
radiation are required, particularly in the region near $z = 1$.

The renormalization/factorization/fragmentation scale dependence of our
isolated cross section is illustrated in Fig. 10 as a function of 
$p_T^c$.  The cuts are those of Fig. 3 for cone size $R=0.7$.
All three scales are varied simultaneously, $\mu=n p^\gamma_T$. The 
dependence on scale is 
negligible in the region $15\leq p_T^c\leq 45$ GeV, whereas outside this
region we see some dependence. As remarked earlier in our discussion of
Fig. 3, both two-body and three-body processes contribute to the cross section
in the region $15\leq p_T^c\leq 45$.  In this region, the next-to-leading order
process is complete in the sense that there are both $O(\alpha_s)$ and
$O(\alpha^2_s)$ contributions.  The factorization scales in the hard
subprocess cross section compensate and cancel those in the structure 
and fragmentation functions. In the regions $p_T^c\leq 15$ GeV and
$p_T^c\geq 45$ GeV, only three-body processes contribute.  There
are no factorization scales in the hard subprocess cross sections,
except for photon fragmentation scales in the latter region, to
compensate the scale dependence of the structure functions.      
The absence  of compensating terms results in the observed scale dependence in 
these regions, particularly in the region of small $p_T$.

A useful measure of the importance of next-to-leading order contributions
is the ``$K$-factor", defined as the ratio of the full cross section through
next-to-leading order to the full leading-order cross section, with 
fragmentation included.  We provide values of $K$ in Fig. 11 as a 
function of $p_T^{\gamma}$ for two different
sets of kinematic selections.  Both curves represent isolated photon
cross sections with $R=0.7$. The solid curve is the $K$-factor 
appropriate to the selections of our Fig. 5. A value of $K$ near
$1.5$ is in agreement with preliminary experimental indications\cite{cdf}. 
The dashed curve in Fig. 11 is the $K$-factor for the cross section with 
the cuts specified in Fig. 4.  

\section{Summary and Discussion}

In this paper we present the results of a calculation of
the inclusive production of a prompt photon in association with a heavy quark at
large values of transverse momentum.  This analysis is done at
next-to-leading order in perturbative QCD.  We employ a combination of 
analytic and Monte Carlo integration methods in which infrared and 
collinear singularities of the next-to-leading order matrix elements are
handled properly.  Our results agree quantitatively with those we obtained using
purely analytic methods\cite{bergergordon}, as they should, but the 
combination of analytic and Monte Carlo methods used in this paper is more
versatile.  We provide differential cross sections in transverse
momenta and rapidity, including
photon isolation restrictions, that should facilitate contact with experimental
results at hadron collider energies.  We show that the study of
two-particle inclusive distributions, with specification
of the momentum variables of both the final prompt photon and the final heavy
quark, tests correlations inherent in the QCD matrix elements\cite{elbcor} 
and should provide a means for measuring the charm quark density in the 
nucleon\cite{bergergordon}.  
Our results are presented in terms of the transverse momentum of the charm
quark.  In a typical experiment\cite{cdf}, the momentum
of the quark may be inferred from the momentum of prompt lepton decay products
or the momentum of charm mesons, such as $D^*$'s.  Alternatively, our
distributions in $p^c_{T}$ may be convoluted with charm quark
fragmentation functions, deduced from, e.g., $e^+e^-$ annihilation
data, to provide distributions for the prompt leptons or $D^*$'s.

\section{Acknowledgments}

The work at Argonne National Laboratory was supported by the US Department of
Energy, Division of High Energy Physics, Contract number W-31-109-ENG-38.
This work was supported in part by Eckerd College.
\pagebreak
 
\appendix

\section{Analytic Expressions}

In order to make this paper reasonably self-contained, we collect in
this appendix all the formulae we use in the calculation. If we label
the momenta for the generic three-body final-state process by
\begin{equation}
p_1+p_2\rightarrow p_3+p_4+p_5 ,
\label{eq:aone}
\end{equation}
where $p_3$ denotes the photon and $p_4$ denotes the observed charm quark, we 
can define the Mandelstam invariants
\begin{eqnarray}
\hat{t}&=&(p_1-p_3)^2 \nonumber \\
\hat{u}&=&(p_2-p_3)^2 \nonumber \\
\hat{s}&=&(p_1+p_2)^2.
\label{eq:atwo}
\end{eqnarray}
We express the two-body final-state cross sections in terms of the scaled
variable $v$, where
\begin{equation}
v=1+\frac{\hat{t}}{\hat{s}}.
\label{eq:athree}
\end{equation}

\subsection{Two-body contributions}

The effective two-body contribution includes the $O(\alpha ^2_s)$ virtual
gluon-exchange loop
contributions and the soft and/or collinear parts of the three-body
contributions (this remark applies to initial-state collinear contributions
only, as explained later).  After all soft pole singularities are 
cancelled and all collinear pole singularities are facto,
the effective two-body contribution is expressed as
\begin{eqnarray}
\sigma_{2,body}(A+B\rightarrow\gamma+c+X)&=& \int dv dx_1 dx_2\left[
\frac{d\sigma_{coll}^{cg\rightarrow \gamma c}}{dv} \right. \nonumber \\
&+&\left. f^A_g(x_1,M^2)f^B_c(x_2,M^2)\frac{d\sigma^{HO}}{dv}(cg\rightarrow
\gamma c)\right],
\label{eq:afour}
\end{eqnarray}
plus terms in which the beam and target are interchanged. In this section we 
use subscript $c$ to refer to the charm (or anti-charm) quark. 

We define 
\begin{equation}
T_{cg}=\frac{2-2v+v^2}{1-v} ,
\end{equation}
\label{eq:afive}
and $v_1=1-v$.  In Eq.~(\ref{eq:afour}),
\begin{eqnarray}
\frac{d\sigma^{HO}}{dv}(cg\rightarrow \gamma c)&=&\frac{\pi\alpha_{em}
\alpha_s e_c^2}{\hat{s}N_C}
\left( T_{cg}+\frac{\alpha_s}{2\pi}
\left[\frac{1}{2}\left(\frac{N_F}{3}T_{cg} -14 C_F T_{cg} + 2 N_C \ln v_1 
+ 4 (2 C_F + N_C) \right. \right. \right. \nonumber \\
&\times & \left. \left. \left. T_{cg} \ln^2\delta_s -\frac{2}{3} N_F T_{cg} 
\ln \frac{\hat{s}}{M^2} +
   C_F T_{cg} (3 + 4 \ln\delta_s) \ln \frac{\hat{s}}{M^2} + \right. \right.
\right. \nonumber \\
& & \left. \left. \left.   C_F T_{cg} (3 + 4 \ln\delta_s) \ln \frac{\hat{s}}{M''^2}
+  \frac{1}{3} N_C T_{cg} (11 + 12 \ln \delta_s)\ln \frac{\hat{s}}{M^2}+ 
\right. \right. \right. \nonumber \\
& & \left. \left. \left. \frac{1}{3}(11 N_C - 2 N_F) T_{cg}\ln \frac{\hat{s}}{\mu^2} + 
4 (2 C_F - N_C) \ln v + 
  4 T_{cg} \ln \delta_s (N_C \ln v_1 +\right.  \right. \right. \nonumber \\
& &\left. \left. \left.
2 C_F \ln v - N_C \ln v) + N_C\ln^2 v_1 (1 + v) +
  (2 C_F - N_C) \ln^2 v\right.  \right. \right. \nonumber \\
&\times &\left. \left. \left. \frac{(2 - 2 v + 3 v^2)}{v_1} 
+ 2 C_F \ln v_1 \frac{(1 + 2 v)}{v_1} - 
  N_C \pi^2 \frac{(-1 - 2 v + 4 v^2)}{3 v_1}\right. \right. \right. \nonumber\\
&+& \left. \left. \left.
  2 C_F \pi^2 \frac{(1 - 4 v + 5 v^2)}{3 v_1} 
+ 2 C_F \ln^2 v_1 \frac{(v^2 + v_1^2)}{v_1} - 
  2 (2 C_F - N_C) \ln v_1\right.  \right. \right. \nonumber \\
&\times & \left. \left. \left. \ln v \frac{(v^2 + v_1^2)}{v_1} 
+  2 (2 C_F - N_C) T_{cg} {\rm Li}_2(1 - v) + 
  2 N_C T_{cg}{\rm Li}_2(v)\right) \right] \right).
\label{eq:asix}
\end{eqnarray}
The scales $M$ and $M''$ are the factorization and fragmentation scales,
respectively, on the initial
parton and final-state charm quark legs, and $\mu$ is the 
renormalization scale. $C_F=4/3$ is the
quark-gluon vertex color factor, $N_C=3$ is the number of colors, $e_c$
is the fractional charge of the charm quark, $\delta_s$ and $\delta_c$
are the soft and collinear cutoff parameters defined in Sec. II,
${\rm Li}_2(x)$ is the dilogarithm function, and $\alpha_{em}$ is the
electromagnetic coupling constant. 

The remnants of the factorization of the hard collinear singularities
are
\begin{eqnarray}
\frac{d\sigma^{cg\rightarrow\gamma c}_{coll}}{dv}&=&\frac{\alpha_{em}
\alpha_s^2e_c^2}{2\hat{s}}\frac{1}{N_C}T_{cg}\nonumber \\
&\times&\left[ f^A_g(x_1,M^2)\left(
\int^{1-\delta_s}_{x_2}\frac{dz}{z}f^B_c\left(\frac{x_2}{z},M^2\right)
\tilde{P}_{qq}(z)+\int^1_{x_2}\frac{dz}{z} f^B_g\left( \frac{x_2}{z},M^2 \right)
\tilde{P}_{qg}(z)\right)\right.  \nonumber \\
&+&\left. 
f^B_c(x_2,M^2)\left(
\int^{1-\delta_s}_{x_1}\frac{dz}{z}f^A_g\left(\frac{x_1}{z},M^2\right)
\tilde{P}_{gg}(z)+\int^1_{x_1} f^A_q\left( \frac{x_1}{z},M^2 \right)
\tilde{P}_{gq}(z)\right) \right].
\label{eq:aseven}
\end{eqnarray}
The last distribution, $f^A_q(x_1,M^2)$, in Eq.~(\ref{eq:aseven}) implies a sum
over the flavors of quarks from the $c c$, $cq$, and
$c\bar{q}$ initial states. The remaining two processes, $c\bar{c}$ and
$q\bar{q}$, do not have initial state collinear singularities and
thus do not contribute to this part of the cross section. Equation 
(\ref{eq:aseven}) differs slightly from that given in Ref.\cite{baer1}, where
the upper limit of integration was always taken to be $1-\delta_s$.
In principle, this is incorrect since it implies that
processes other than the $cg$ initiated process may have soft
singularities, but the error introduced by that approximation is very small
for small $\delta_s$.  We use the correct expression in our calculation.

The splitting functions $\tilde{P}_{ij}$, listed in the appendix of
Ref.\cite{baer1}, are included here for completeness.
\begin{equation}
\tilde{P}_{ij}(z)=P_{ij}(z)\ln\left(\frac{1-z}{z}\delta_c
\frac{\hat{s}}{M^2}\right)-P'_{ij}(z).
\label{eq:aeight}
\end{equation}
The functions $P_{ij}(z)$ are the usual Altarelli-Parisi splitting functions in
$4-2\epsilon$ dimensions and are 
\begin{eqnarray}
P_{qq}(z,\epsilon)&=&C_F\left[\frac{1+z^2}{1-z}-\epsilon(1-z)\right]\nonumber \\
P_{qg}(z,\epsilon)&=&\frac{1}{2(1-\epsilon)}\left[z^2+(1-z)^2-\epsilon\right]\nonumber \\
P_{gg}(z,\epsilon)&=&2N_C\left[\frac{z}{1-z}+\frac{1-z}{z}+z(1-z) \right]\nonumber \\
P_{gq}(z,\epsilon)&=&C_F\left[\frac{1+(1-z)^2}{z}-\epsilon z\right].
\label{eq:anine}
\end{eqnarray}
The functions $P'_{ij}(z)$ are defined by the relation
\begin{equation}
\tilde{P}_{ij}(z,\epsilon)=P_{ij}(z)+\epsilon P'_{ij}(z).
\label{eq:aten}
\end{equation}

\subsection{Pseudo-Two-Body Contributions}

Since we are interested in distributions in the kinematic variables of
two final-state partons, the photon and the $c$ quark, we can define
variables that depend on the momenta of both.  An example is
the variable $z$, defined in Eq.~(\ref{eq:zdef}).  Whenever there is a third 
parton in the final state, the distribution in $z$ (or in other analogous 
variables) will differ from a delta-function 
when the third parton carries a finite momentum,
even if it is collinear to one of the other final partons. For this reason we
designate as ``pseudo-two-body contributions" those for which the third parton
is collinear to either the final photon or the charm quark. These
contributions are expressed, respectively, by the equations 
\begin{eqnarray}
\sigma_{\gamma/coll}&=&\sum_{abcq}\int
f^A_a(x_1,M^2)f^B_b(x_2,M^2)\left(\frac{\alpha_{em}}{2\pi}\right) \left[
P_{\gamma q}(z)\ln\left[z(1-z)\delta_c\frac{\hat{s}}{M'^2}\right]-
P'_{\gamma/q}(z)
\right]  \nonumber \\
&\times &\frac{d\hat{\sigma}}{dv}
(ab\rightarrow cq)dx_1 dx_2 dz dv,
\label{eq:aeleven}
\end{eqnarray}
and
\begin{equation}
\sigma_{c/coll}=\sum_{abd}\int
f^A_a(x_1,M^2)f^B_b(x_2,M^2)\tilde{P}_{cd}(z,M''^2)\frac{d\hat{\sigma}}{dv}
(ab\rightarrow \gamma d)dx_1 dx_2 dz dv.
\label{eq:atwelve}
\end{equation}
The functions $P_{\gamma/q}(z)$ and $P'_{\gamma/q}(z)$ are the quark-to-photon
splitting function and $O(\epsilon)$ piece, respectively.  They have the same
form as $P_{gq}$, with the color factor replaced by the square of the quark 
charge.  The scale $M'$ is the fragmentation scale for quark fragmentation 
into a photon.
\begin{equation}
\tilde{P}_{cd}(z,M''^2)=P_{cd}(z)\ln\left[z(1-z)\delta_c\frac{s}{M''^2}\right]
-P'_{cd}(z),
\label{eq:athirteen}
\end{equation}
where $P_{cd}(z)$ represents the splitting functions $P_{qg}(z)$ and
$P_{qq}(z)$ of Eq.~(\ref{eq:anine})  along with
the `primed' pieces.  In Eq.~(\ref{eq:aeleven}), $q$ can be a charm or 
anti-charm
quark, or a (anti) quark of any flavor in the case of $c q\rightarrow c q$. 
In Eq.~(\ref{eq:atwelve}), $d$ is either a gluon or (anti) charm
quark. 

These contributions are usually referred to as the remnants of the
factorization of the hard collinear singularities and are regarded as two-body
processes, or as parts of the fragmentation contributions because of their
dependence on the factorization scales.  We prefer to regard
them as pseudo-two-body contributions.  When we examine either the
charm quark or photon momentum distributions, these contributions populate
the same regions of phase space as the other three-body contributions in 
Eq.~(\ref{eq:aseventeen}), unlike the effective two-body contributions.
The pseudo-two-body contributions are usually negative in overall sign due to
the large logarithms of the cut-off parameters, as are the two-body
contributions discussed above. 

\subsection{Photon Fragmentation Contributions}

As mentioned in Sec.~III, we include the quark-to-photon and gluon-to-photon
fragmentation contributions at leading order only.  We convolute the 
$2\rightarrow 2$ hard scattering subprocess cross sections for the processes
listed in Eq.~(\ref{eq:fragproc}) with photon fragmentation functions
$D_{\gamma/i}(z,M'^2)$; $M'$ is the fragmentation scale, the same
scale at which we subtract the collinear singularities on the photon
leg of the three-body processes. The expression for the cross section is
\begin{equation}
\sigma_{frag/\gamma}=\sum_{abj}\int
f^A_a(x_1,M^2)f^B_b(x_2,M^2)D_{\gamma/j}(z,M'^2)\frac{d\hat{\sigma}}{dv}
(ab\rightarrow jc)dx_1 dx_2 dz dv.
\label{eq:afourteen}
\end{equation}
The matrix elements for the hard subprocess cross section are well known
and can be found, for example, in Ref.\cite{frag1}.

\subsection{Charm Fragmentation Contributions}

In integrating some of the three-body matrix elements over phase space we
encounter configurations in which the charm quark is produced collinearly
with an anti-charm quark or a gluon in the final state.  These situations lead
to a collinear singularity in the massless approximation.  They occur for 
the processes of Eq.~(\ref{eq:11}), (\ref{eq:13}), and (\ref{eq:16}). We
factor these singularities into a fragmentation function $D_{c/i}(z,M''^2)$ for
parton $i$ to produce a charm quark with momentum fraction $z$. The
contributing subprocess cross sections are 
\begin{eqnarray}
\frac{d\hat{\sigma}}{dv}(q\bar{q}\rightarrow \gamma
g)&=&\frac{2C_F}{N_C}\frac{\pi\alpha\alpha_s e_q^2}{s}\left(
\frac{v}{1-v}+\frac{1-v}{v}\right) ;\nonumber \\
\frac{d\hat{\sigma}}{dv}(qg\rightarrow \gamma
q)&=&\frac{\pi\alpha\alpha_s e_q^2}{N_C s}\left(
\frac{1+(1-v)^2}{1-v}\right).
\label{eq:afifteen}
\end{eqnarray}
The physical cross section is given by 
\begin{equation}
\sigma_{frag/c}=\sum_{abd}\int
f^A_a(x_1,M^2)f^B_b(x_2,M^2)D_{c/d}(z,M''^2)\frac{d\hat{\sigma}}{dv}
(ab\rightarrow \gamma d)dx_1 dx_2 dz dv.
\label{eq:asixteen}
\end{equation}

\subsection{Three-body contributions}

The non-collinear three-body final-state contributions are calculated from 
the expression
\begin{equation}
\sigma_{3-body}=\sum_{abd}\int f^A_a(x_1,M^2)
f^B_b(x_2,M^2)d\hat{\sigma}(ab\rightarrow\gamma cd)dx_1dx_2d\Omega,
\label{eq:aseventeen}
\end{equation}  
with $\Omega$ representing the angles and other variables that are integrated 
over. Whenever an invariant $s_{ij}$ or $t_{ij}$ falls into a collinear or 
soft region of phase space, that contribution from the subprocess is excluded.
The three-body contribution shows no dependence on
the factorization scale of the final-state charm or photon legs,
although we have factored collinear singularities at scales $M''$ and
$M'$, respectively, on these legs of the three-body subprocesses. 
However, Eq.~(\ref{eq:aseventeen}) does contain implicit logarithmic 
dependence on the soft and, in particular, the collinear cutoffs discussed 
in Sec.~II.  Both collinear cutoff and factorization scale
dependences are contained in the pseudo-two-body contributions
discussed above.

%\noindent

\noindent
\begin{center}
{\large FIGURE CAPTIONS}
\end{center}
\newcounter{num}
\begin{list}%
{[\arabic{num}]}{\usecounter{num}
    \setlength{\rightmargin}{\leftmargin}}

\item (a) Lowest order Feynman diagrams for $\gamma$ plus $c$ quark 
production; $k_1$ and $k_2$ are the four-vector momenta of the photon and
charm quark. (b) Examples of virtual corrections to the lowest order diagrams.
(c) Examples of next-to-leading order three-body final-state diagrams for 
the $g c$ initial state.  

\item Study of the dependence of the cross section on the Monte Carlo 
cutoff parameters $\delta_s$ and $\delta_c$.  Shown is the cross section 
for $p +\bar{p}\rightarrow \gamma + c + X$ at $\sqrt{s}=1.8$ TeV with 
the transverse momenta of the photon and charm quark restricted to the 
the interval $10 < p_T < 50$ GeV and the rapidities of the photon and
charm quark restricted to the interval $-3.0 < y < 3.0$.

\item Cross section $d\sigma/dp^c_{T}$ as a function of the transverse 
momentum of the charm quark for $p +\bar{p}\rightarrow \gamma + c + X$ at 
$\sqrt{s}=1.8$ TeV.  The transverse momentum of the photon is restricted to
the interval $15 < p^\gamma_T < 45$ GeV, and the rapidities of the photon and
charm quark are restricted to the interval $-1.0 < y < 1.0$.  Four curves are
drawn.  The solid curve shows the cross section with no further restrictions.
The dashed curve indicates the result after the additional selection is made 
that $z > 0.1$; the ratio $z$ is defined in the text.  The dotted and dash-dot
curves display the results after photon isolation restrictions are applied, in
addition to the cut on $z$.

\item Cross section $d\sigma/dp^\gamma_{T}$ as a function of the transverse
momentum of the photon for $p +\bar{p}\rightarrow \gamma + c + X$ at
$\sqrt{s}=1.8$ TeV.  The transverse momentum of the charm quark is restricted to
the interval $15 < p^c_T < 45$ GeV, and the rapidities of the photon and
charm quark are restricted to the interval $-1.0 < y < 1.0$.  Four curves are
drawn.  The solid curve shows the cross section with no further restrictions.
The dashed curve indicates the result after the additional selection is made
that $z > 0.1$; the ratio $z$ is defined in the text.  The dotted and dash-dot
curves display the results after photon isolation restrictions are applied, in
addition to the cut on $z$.

\item Cross section $d\sigma/dp^\gamma_{T}$ as a function of the transverse
momentum of the photon for $p +\bar{p}\rightarrow \gamma + c + X$ at
$\sqrt{s}=1.8$ TeV.  The transverse momentum and rapidity of the charm quark 
are not restricted, but the rapidity of the photon is limited to the 
interval $-0.5 < y < 0.5$.  The ratio $z$ is restricted to the interval 
$0.2 < z < 2.0$.  Three curves are
drawn.  The solid curve shows the cross section with no further restrictions.
The dashed curve indicates the result after photon isolation is imposed, with 
$R$ = 0.7, and the dot-dashed curve is the leading order cross section with
photon isolation imposed, $R$ = 0.7.

\item Cross section $d\sigma/dy^c$ as a function of the rapidity of the charm
quark for $p +\bar{p}\rightarrow \gamma + c + X$ at $\sqrt{s}=1.8$ TeV.  The
photon rapidity is restricted to $-1.0 < y^\gamma < 1.0$, and photon isolation 
is imposed, with $R$ = 0.7.  There is no restriction on $p^c_T$, but the
ratio $z$ is restricted to $z > 0.1$.  Curves are shown for three 
selections on the transverse momentum of the
photon.  The solid, dashed, and dot-dashed curves correspond to the selections
$15 < p^\gamma_T < 45$ GeV, $15 < p^\gamma_T < 25 GeV$, and 
$35 < p^\gamma_T < 45$ GeV, respectively.  For ease of comparison of shapes, the
dot-dashed curve has been multiplied by 10.  
 
\item Cross section $d\sigma/d\Delta y$ as a function of the difference 
$\Delta y = y^\gamma - y^c$ of the rapidities of the photon and charm quark,
for $p +\bar{p}\rightarrow \gamma + c + X$ at $\sqrt{s}=1.8$ TeV.  The
ratio $z$ is restricted to $z > 0.1$, and photon isolation is imposed, with 
$R$ = 0.7.  The transverse momentum of the
photon is selected to be in the interval $15 < p^\gamma_T < 45$ GeV.
In (a), the photon rapidity is restricted to $-1.0 < y^\gamma < 1.0$; in (b)
$1.0 < y^\gamma < 2.0$.  In (b), the dashed curve shows the behavior at 
leading order.  

\item Cross section $d\sigma/dy^c$ as a function of the rapidity of the charm
quark for $p +\bar{p}\rightarrow \gamma + c + X$ at $\sqrt{s}=1.8$ TeV.  Photon 
isolation is imposed, with $R$ = 0.7; $z > 0.1$; and $15 < p^\gamma_T < 45$ GeV.
The solid curve shows the result when the photon rapidity is restricted to 
$1.0 < y^\gamma < 2.0$, and the dashed curve displays the result for 
$2.0 < y^\gamma < 3.0$.

\item Cross section $d\sigma/dp^\gamma_{T} dy^\gamma dz$ as a function of the 
ratio $z$ for $p +\bar{p}\rightarrow \gamma + c + X$ at $\sqrt{s}=1.8$ TeV.  
The transverse momentum and rapidity of the photon are averaged over the 
intervals $14 < p^\gamma_T < 16$ GeV and $-0.5 < y^\gamma < 0.5$.  
In (a), we illustrate the effects of
photon isolation by comparing the distributions in $z$ with and without the
isolation restriction. The solid histograms
in (b) and (c) show the results of our calculation for our standard Monte Carlo 
integration cutoff parameters $\delta_s = 0.01$ and $\delta_c = 0.001$.  In
(b) and (c), photon isolation is imposed with $R$ = 0.7.  In (b), we display 
the dependence of the final cross section on the Monte Carlo integration cutoff
parameter $\delta_s$, having fixed $\delta_c = 0.001$.  In (c), we show the
dependence of the final cross section on the cutoff parameter $\delta_c$,
having fixed $\delta_s = 0.01$.  

\item The renormalization/factorization scale ($\mu$) dependence of the cross 
section is displayed.  Shown is the transverse momentum dependence of 
$d\sigma/dp^c_{T}$ for three values of $\mu/p^\gamma_{T}$: 0.5, 1.0, and 2.
The transverse momentum of the photon is restricted to the interval 
$15 < p^\gamma_T < 45$ GeV, and the rapidities of the photon and charm quark 
are restricted to the interval $-1.0 < y < 1.0$.

\item The K factor defined in the text is shown as a function of 
$p^\gamma_{T}$. Photons are isolated with $R$ = 0.7.  The solid line 
corresponds to selections analogous to those of our analytic paper [Ref. 1]:
$-0.5 < y^\gamma < 0.5$ and $0.2 < z < 2.0$, but with no restrictions on 
$y^c$ or $p^c_T$.  The dashed line shows how results change when the
transverse momentum of the charm quark is restricted to the interval
$15 < p^c_T < 45$ GeV with, in addition, $z > 0.1$ and the rapidities of the
photon and charm quark limited to the interval $-1.0 < y < 1.0$.

\end{list}

\begin{references}

\bibitem{bergergordon} E. L. Berger and L. E. Gordon, Argonne report 
ANL-HEP-PR-9536 (hep-ph/9512343), submitted to Phys. Rev. {\bf D},
and references therein.
\bibitem{cdf} CDF Collaboration, R. Blair {\it{et al}}, 
Proceedings of the 10th Topical Workshop on Proton-Antiproton Collider 
Physics, May, 1995 (AIP Conference Proceedings 357), edited by R. Raja and
J. Yoh (AIP Press, N.Y., 1996), pp 557-567.
\bibitem{field} E. L. Berger, E. Braaten, and R. D. Field, Nucl. Phys.
 {\bf B239}, 52 (1984).
\bibitem{frag1}  D.W. Duke and J. F. Owens, Phys. Rev. {\bf D26}, 1600 (1982);
J. F. Owens, Rev. Mod. Phys. {\bf 59}, 465 (1987).
\bibitem{frag2} P. Aurenche {\it{et al}}, Nucl. Phys. {\bf B399}, 34 (1993).
\bibitem{frag3} M. Gl\"{u}ck, E. Reya, and A. Vogt, Phys. Rev. {\bf D48}, 116
(1993).  
\bibitem{berqiu} E. L. Berger and J.-W. Qiu, Phys. Lett. {\bf B248},
371 (1990) and Phys. Rev. {\bf D44}, 2002 (1991); E. L. Berger, X. Guo, and
J.-W. Qiu, Argonne report ANL-HEP-PR-9588 (hep-ph/9512281) to be published in
Phys. Rev. Lett.
\bibitem{baer1}
 H.~Baer, J.~Ohnemus, and J.~F.~Owens, Phys. Rev. {\bf D42}, 61 (1990).
\bibitem{bailey1}
 B.~Bailey, J.~Ohnemus, and J.~F.~Owens, Phys. Rev. {\bf D46}, 2018 (1992).
\bibitem{baer2} 
 H. Baer and H. Reno, Phys. Rev. {\bf D43}, 2892 (1991).
\bibitem{ohnemus}
J. Ohnemus and J. F. Owens, Phys. Rev. {\bf D43}, 3626 (1991); J. Ohnemus,
Phys. Rev. {\bf D44}, 1403 (1991); J. Ohnemus, Phys. Rev. {\bf D44}, 
3477 (1991).
\bibitem{correl}M. Mangano, P. Nason, and G. Ridolfi, Nucl.Phys. {\bf B373}, 295
(1992).  
\bibitem{gorvogel} L.~E.~Gordon and W.~Vogelsang, Phys. Rev. {\bf D48}, 3136 
(1993) and {\bf D50}, 1901 (1994);
 M.~Gl\"{u}ck, L.~E.~Gordon, E.~Reya, and W.~Vogelsang, Phys. Rev. Lett.
 {\bf 73}, 388 (1994).
\bibitem{altar} G. Altarelli and G. Parisi, Nucl. Phys.
{\bf B126}, 298 (1977). 
\bibitem{parden2} H. L. Lai {\it et al}, Phys. Rev. {\bf D51}, 4763 (1995).
\bibitem{elbcor} E.~L.~Berger, Phys. Rev. {\bf D37}, 1810 (1988).
\end{references}
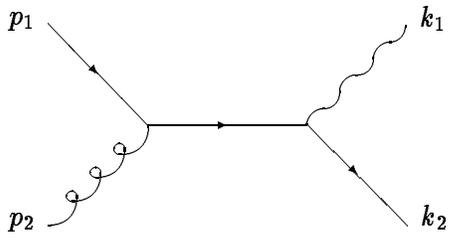
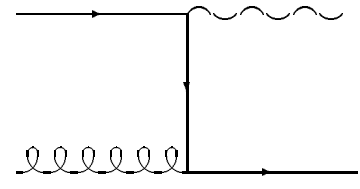
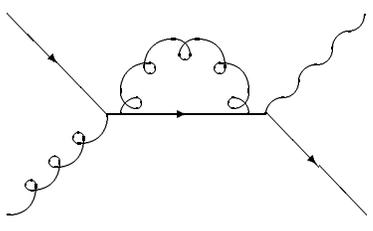
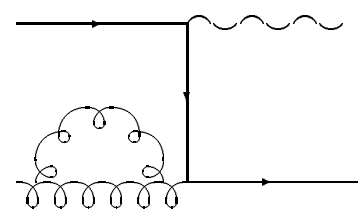
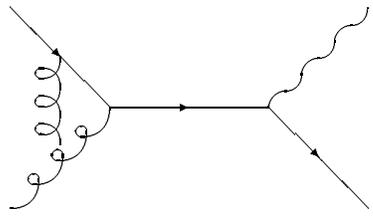
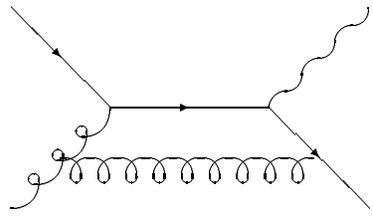
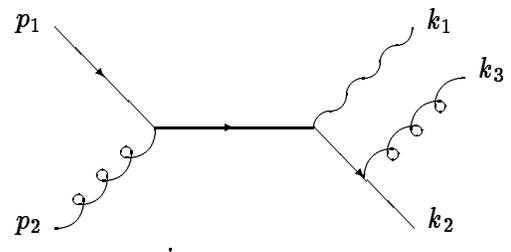
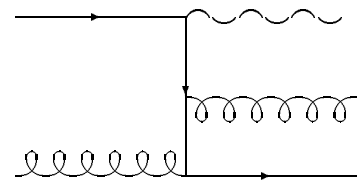
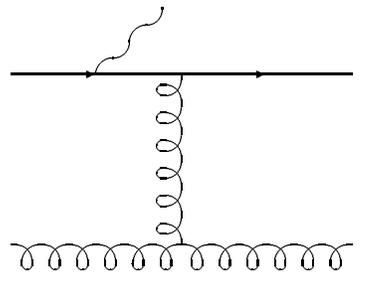
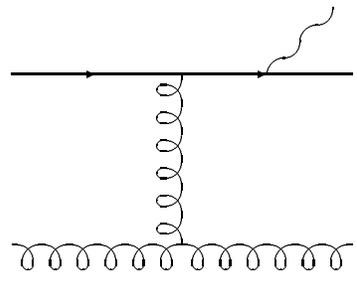
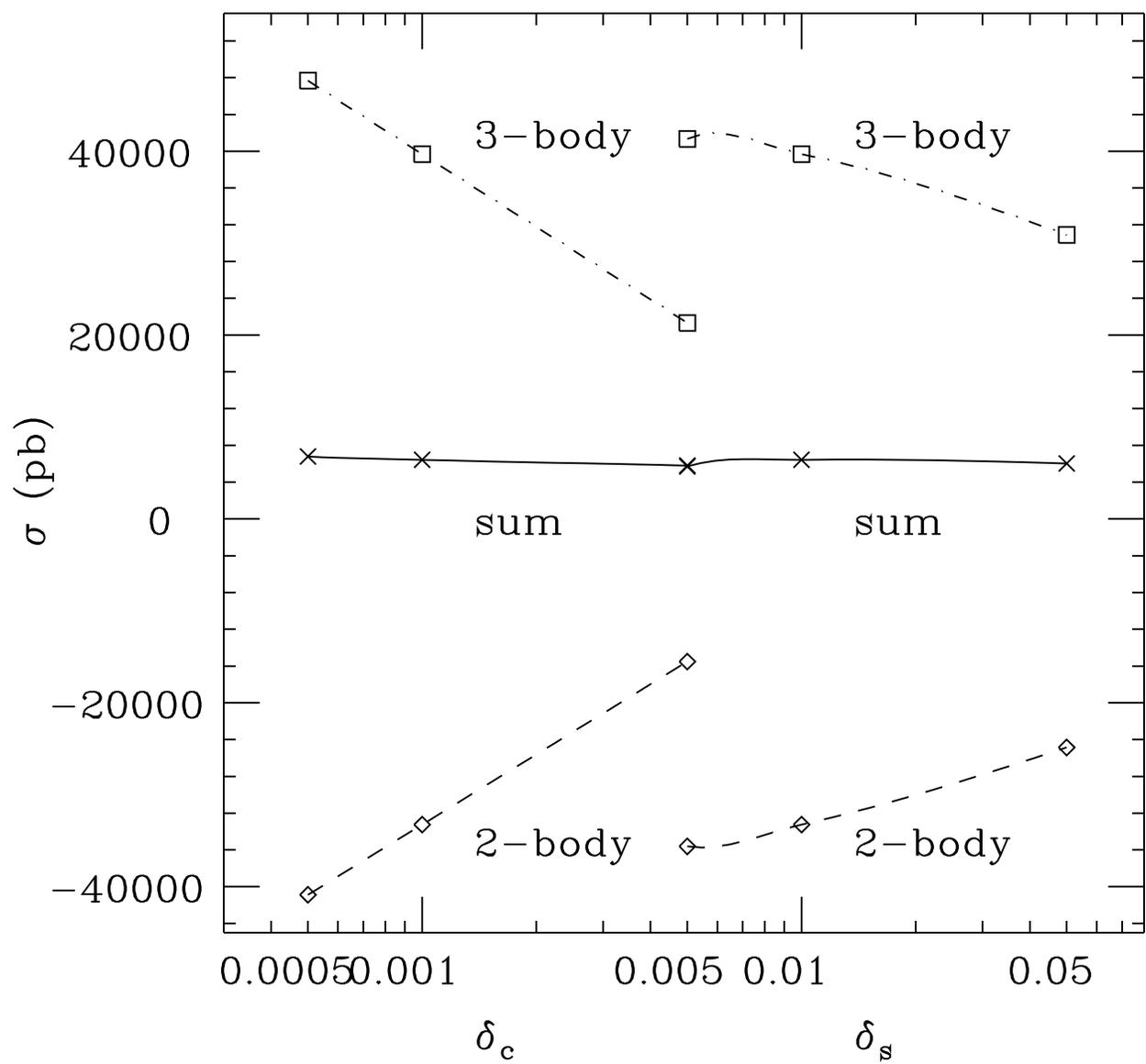
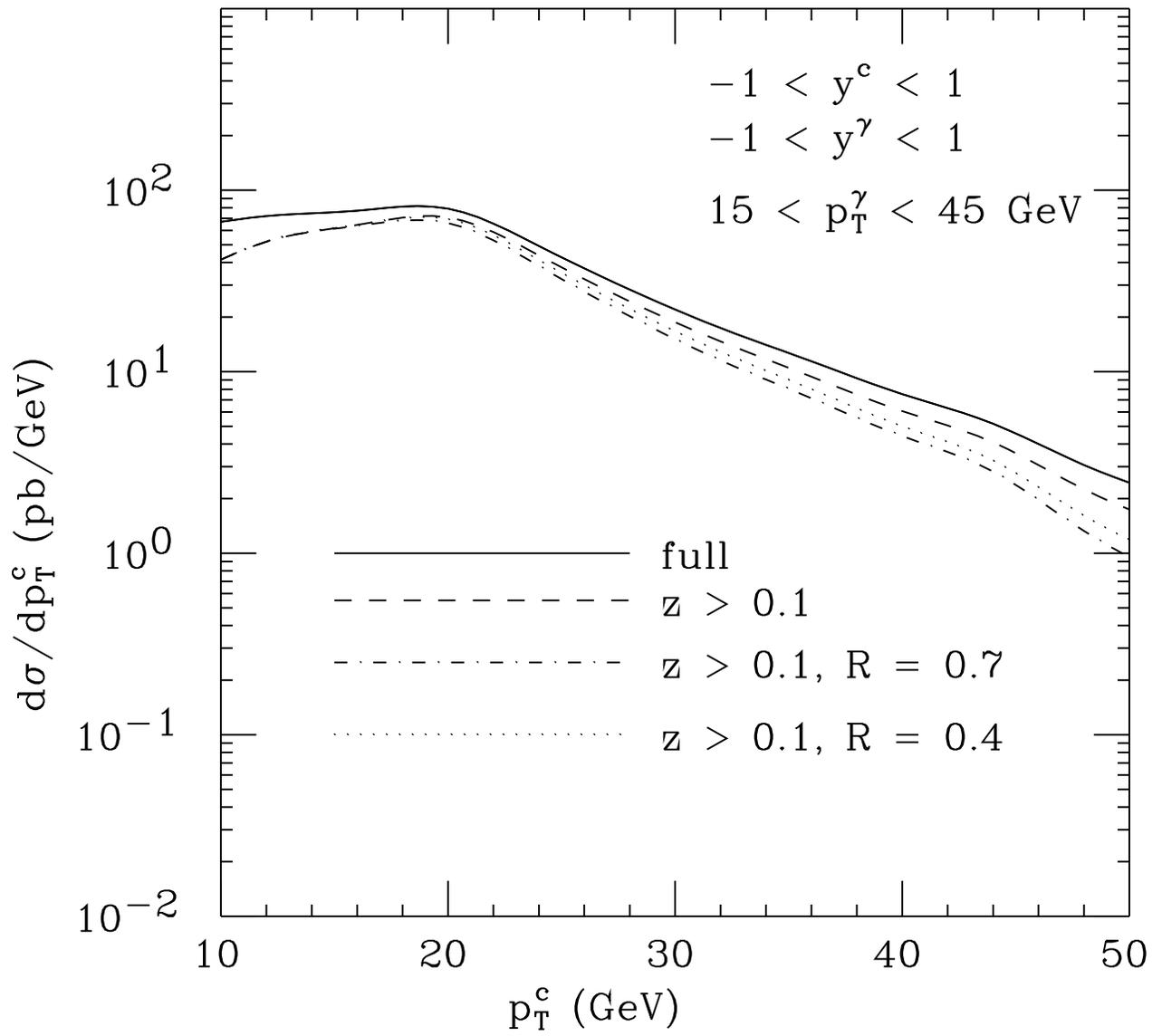
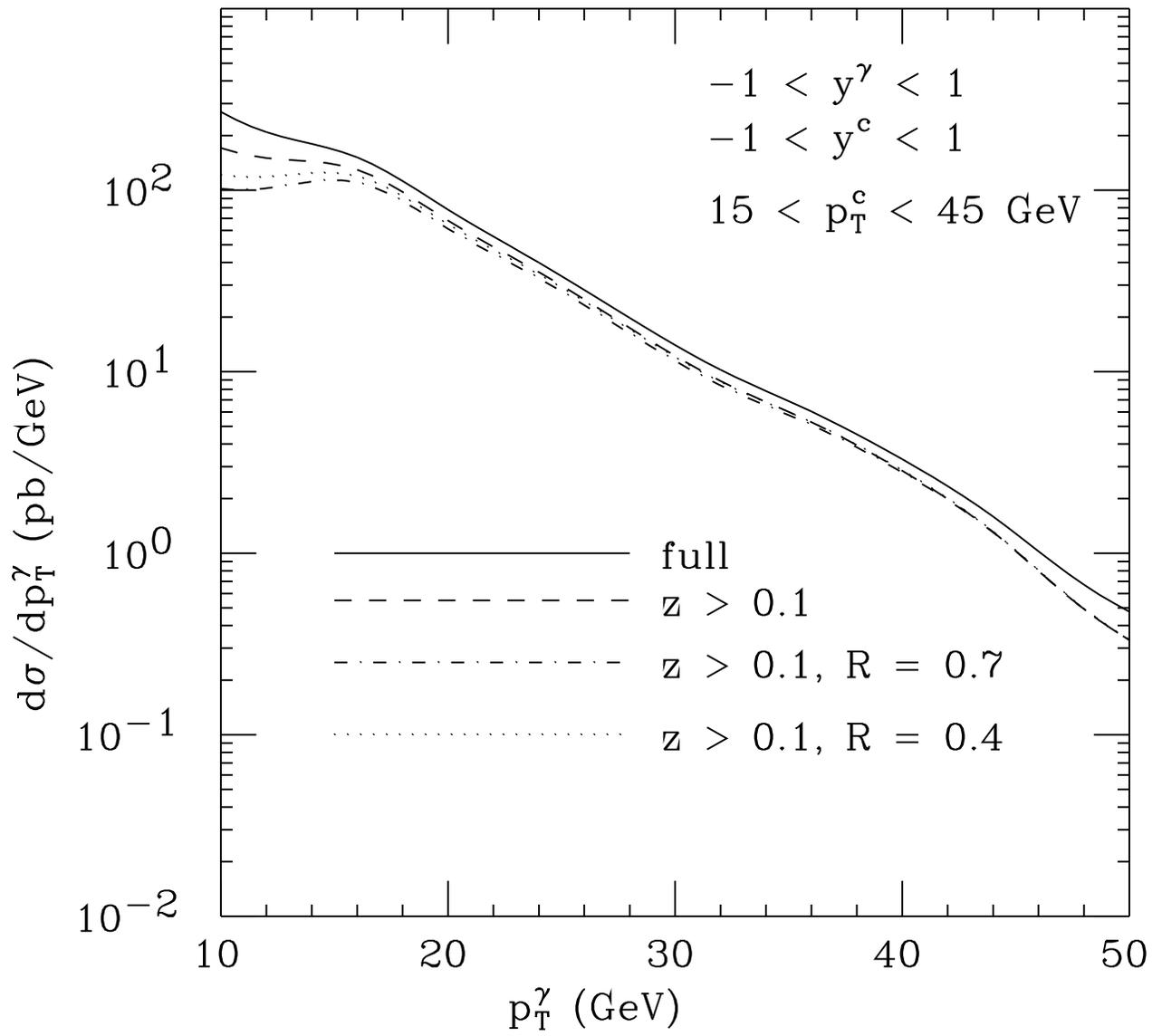
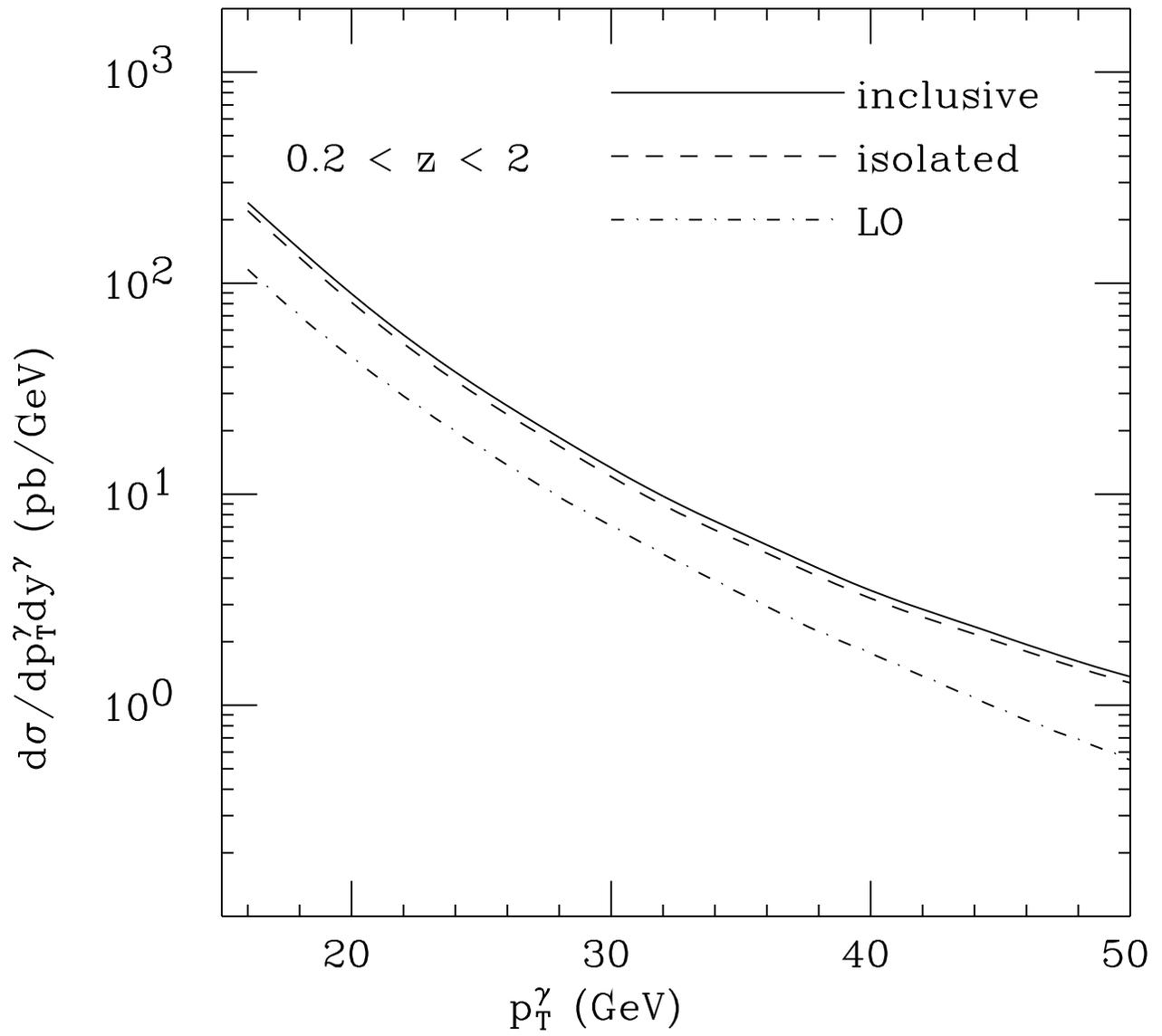
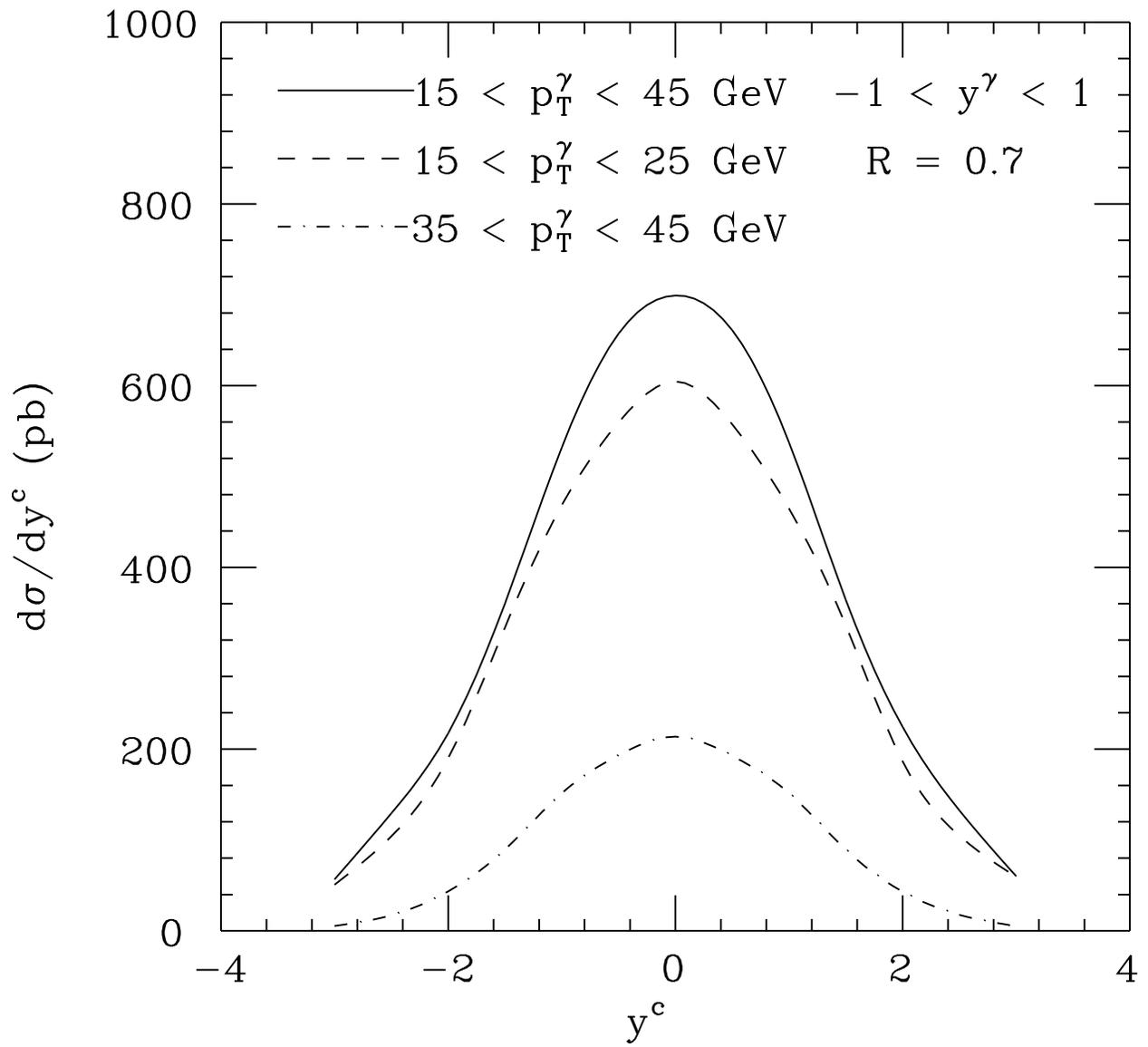
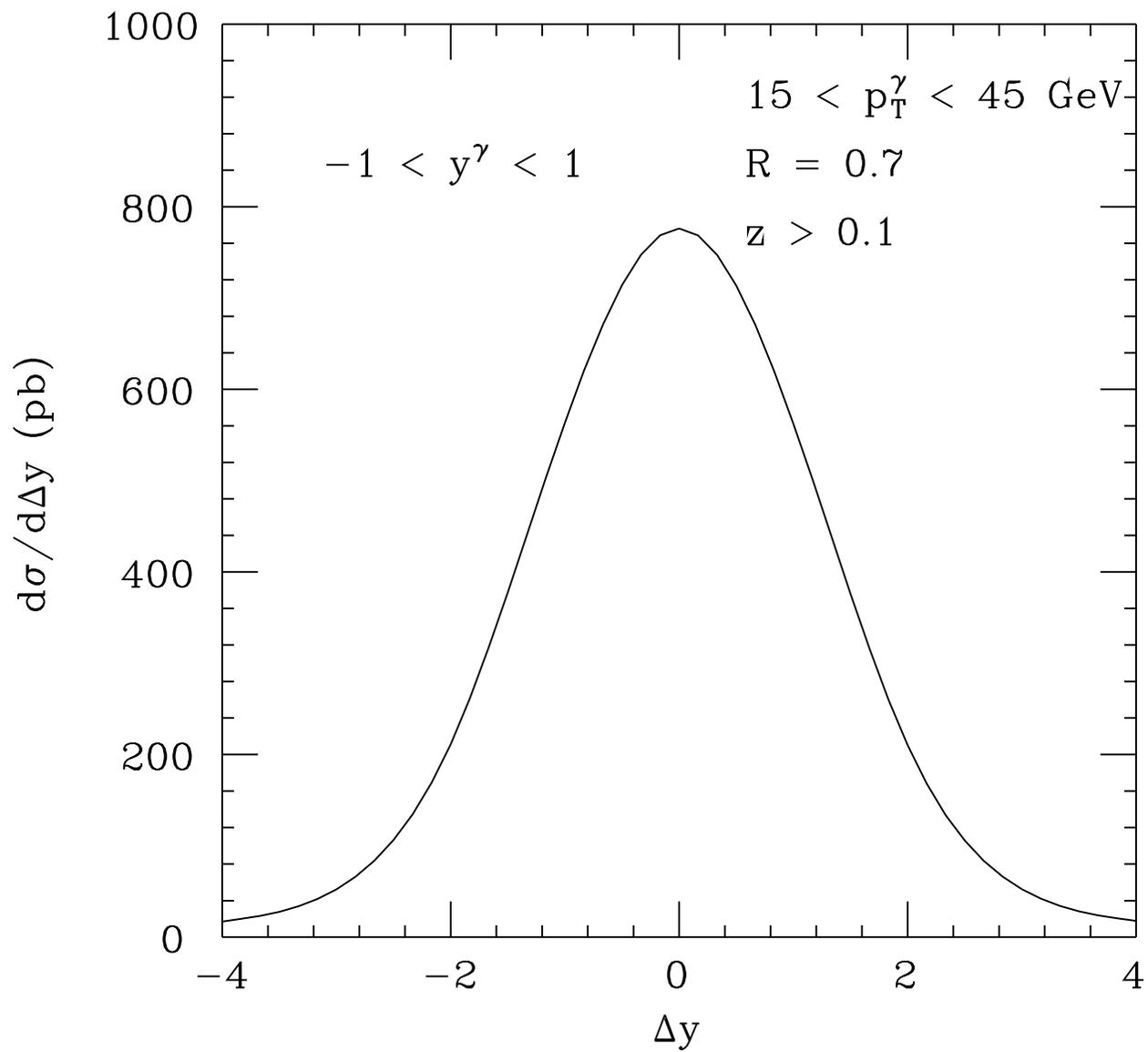
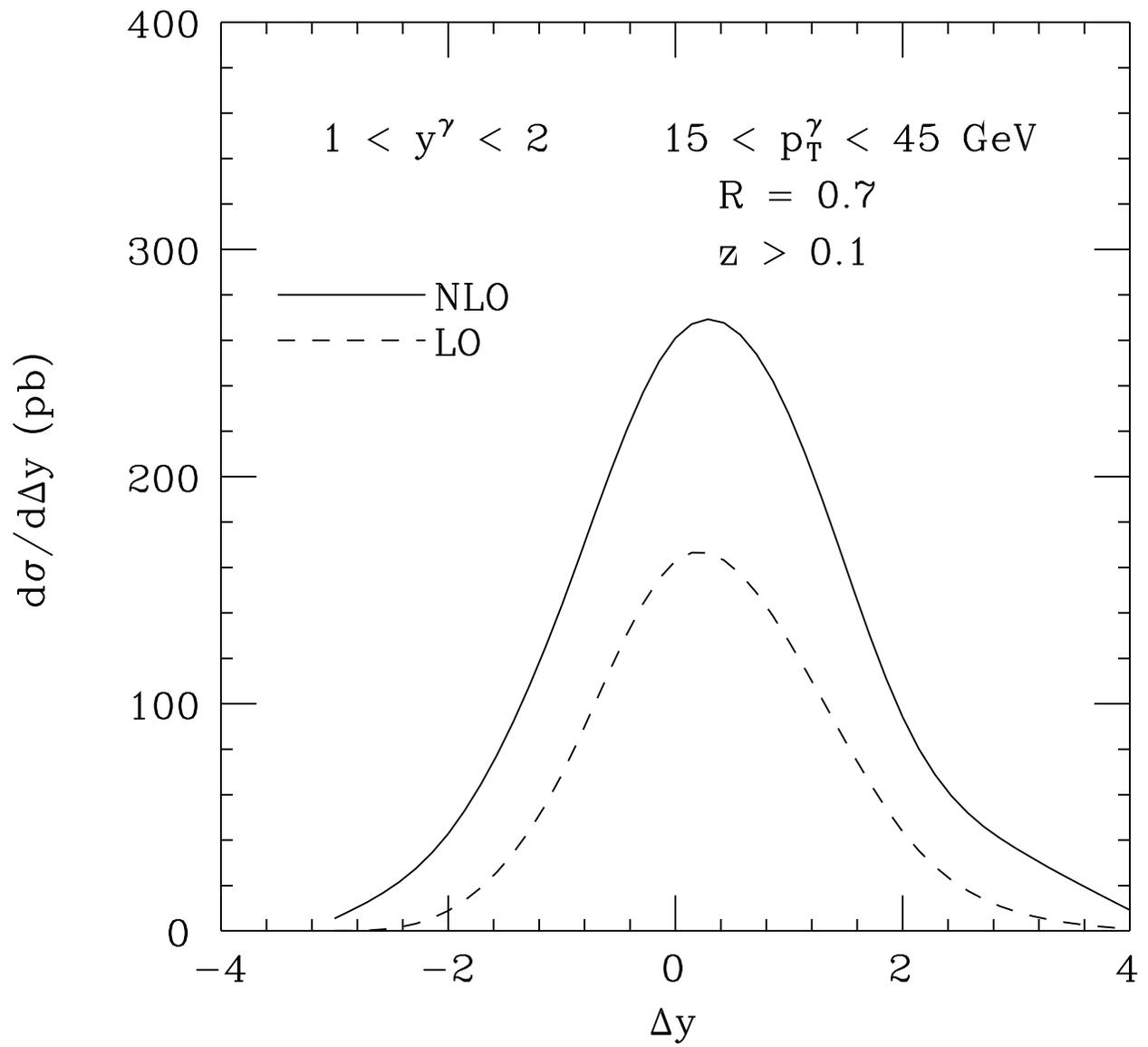
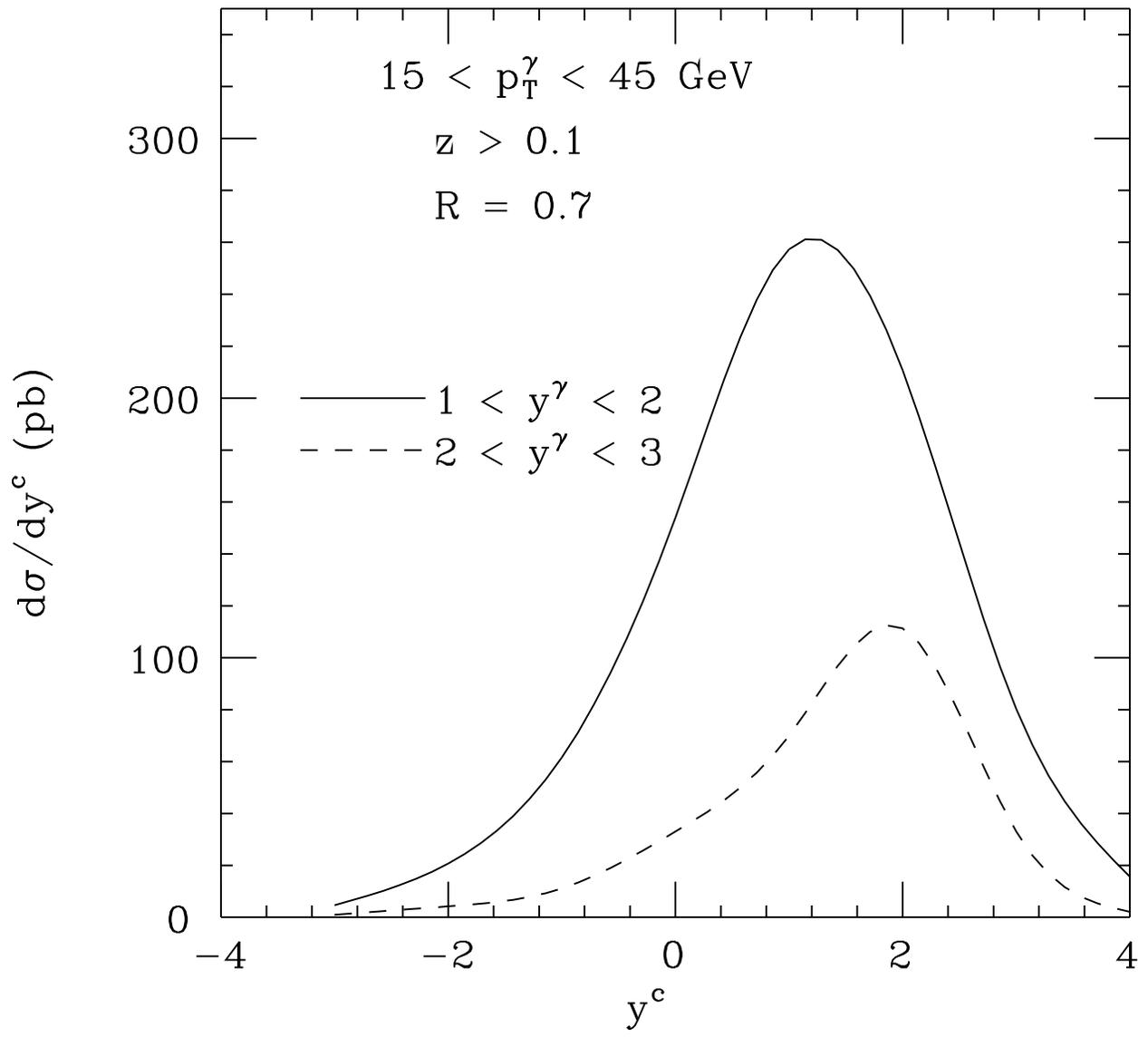
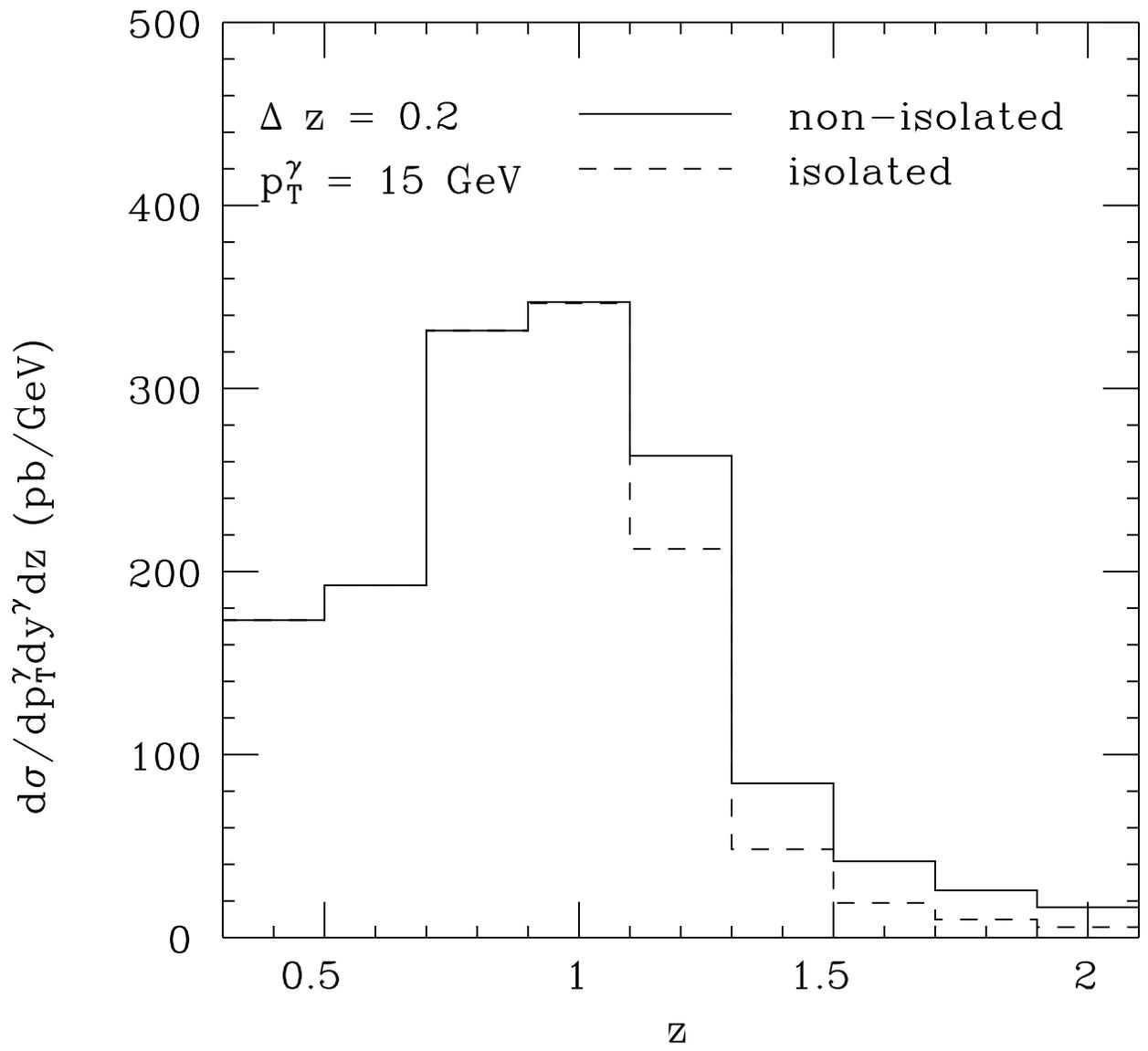
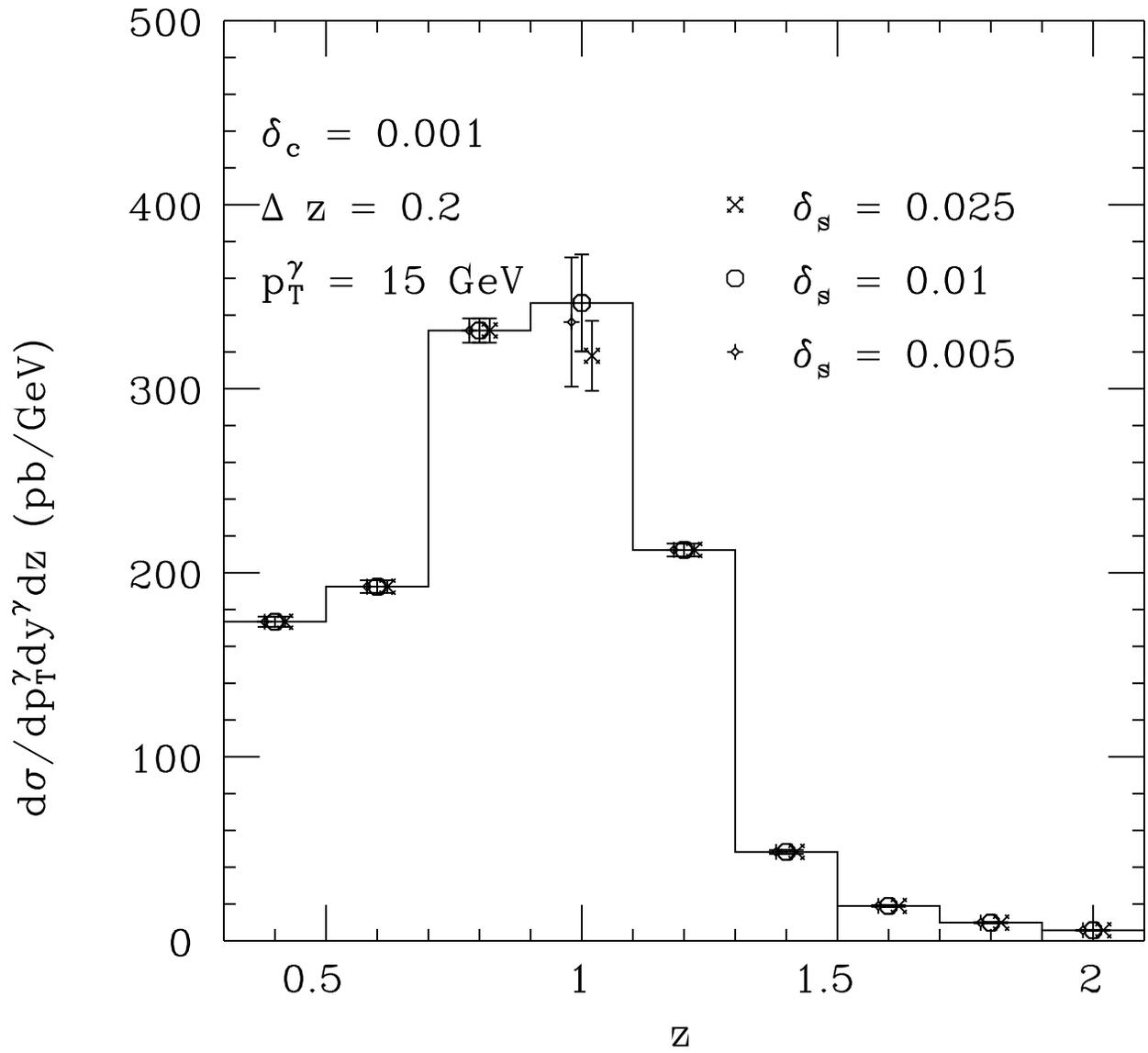
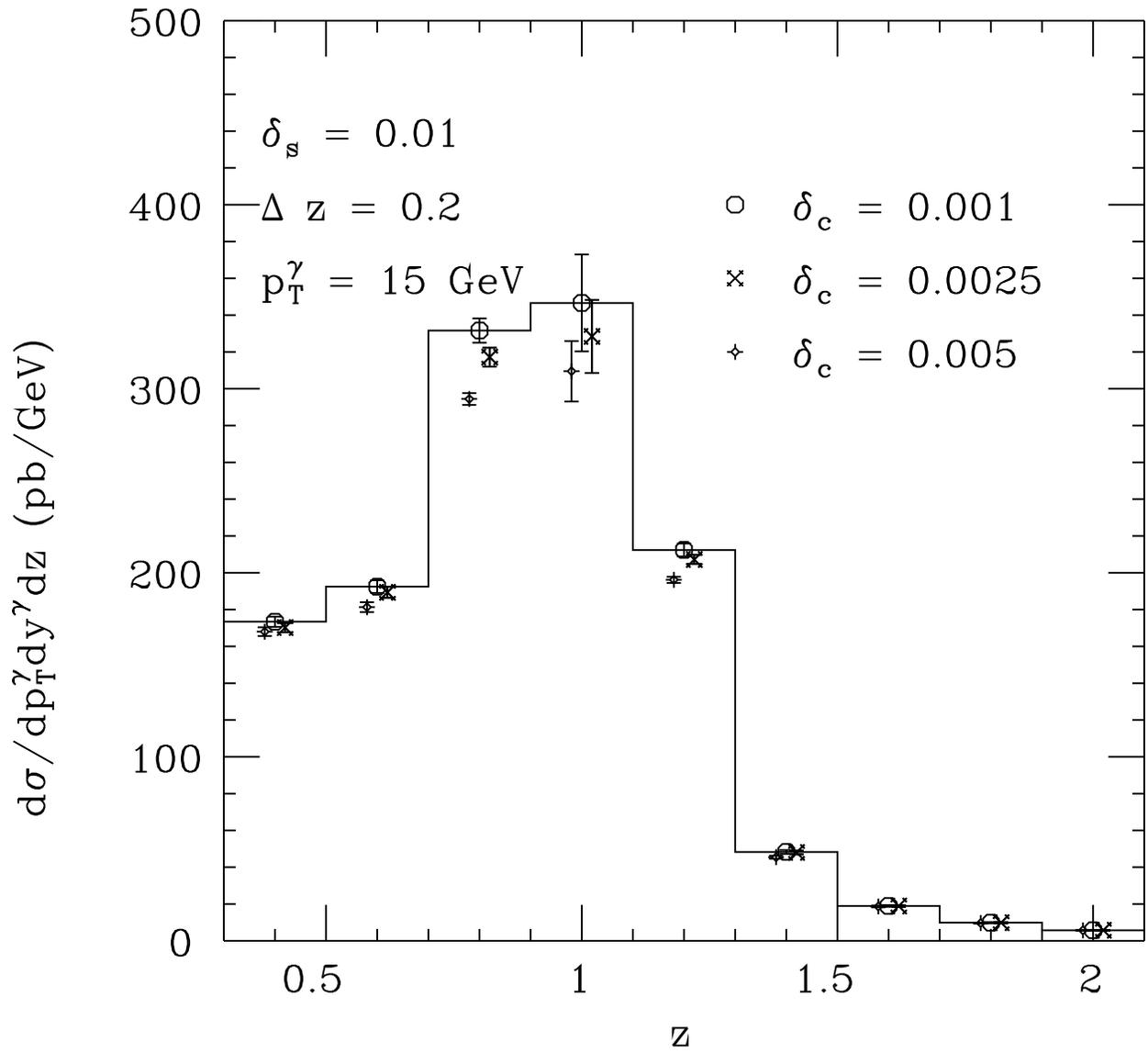
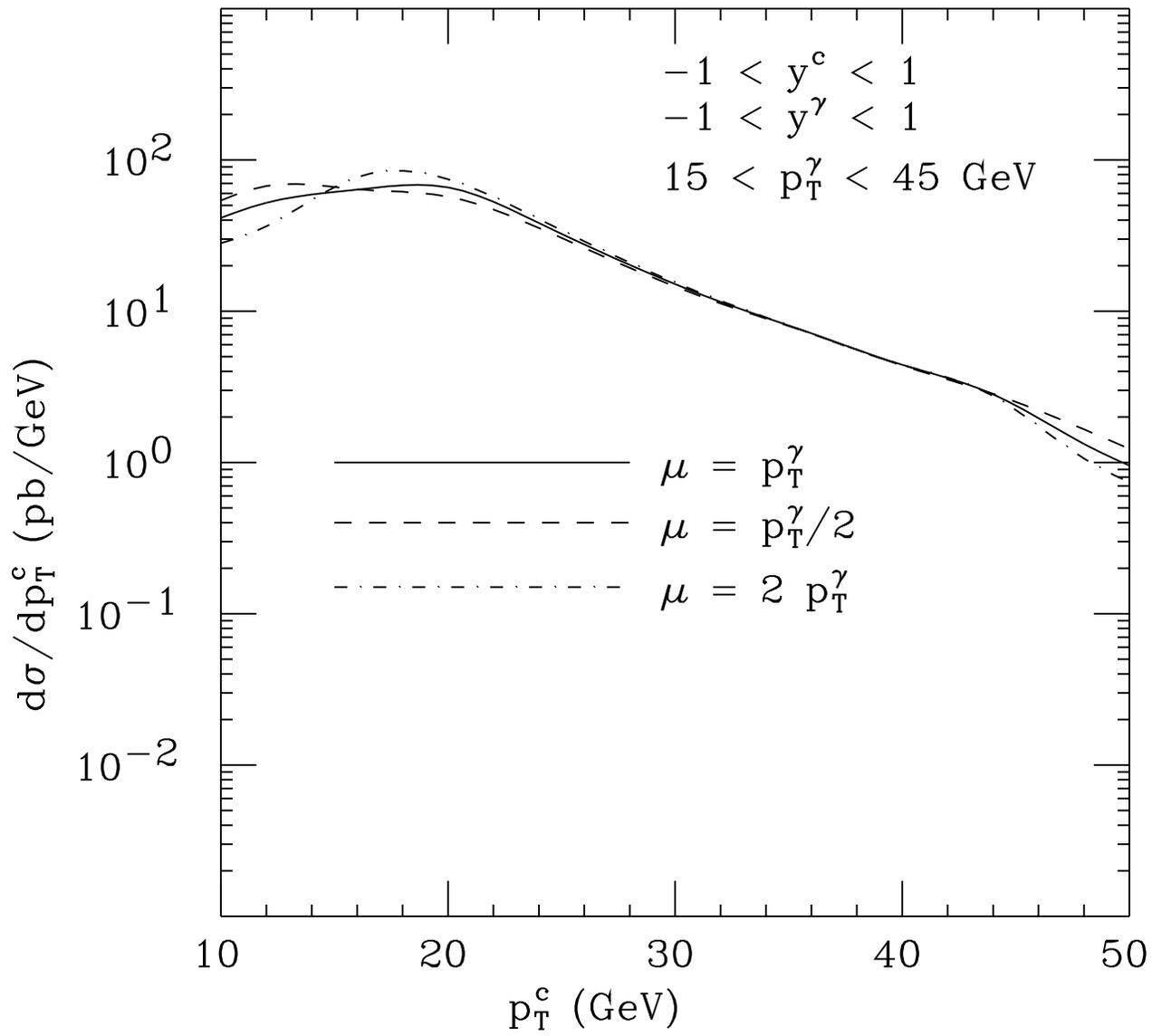
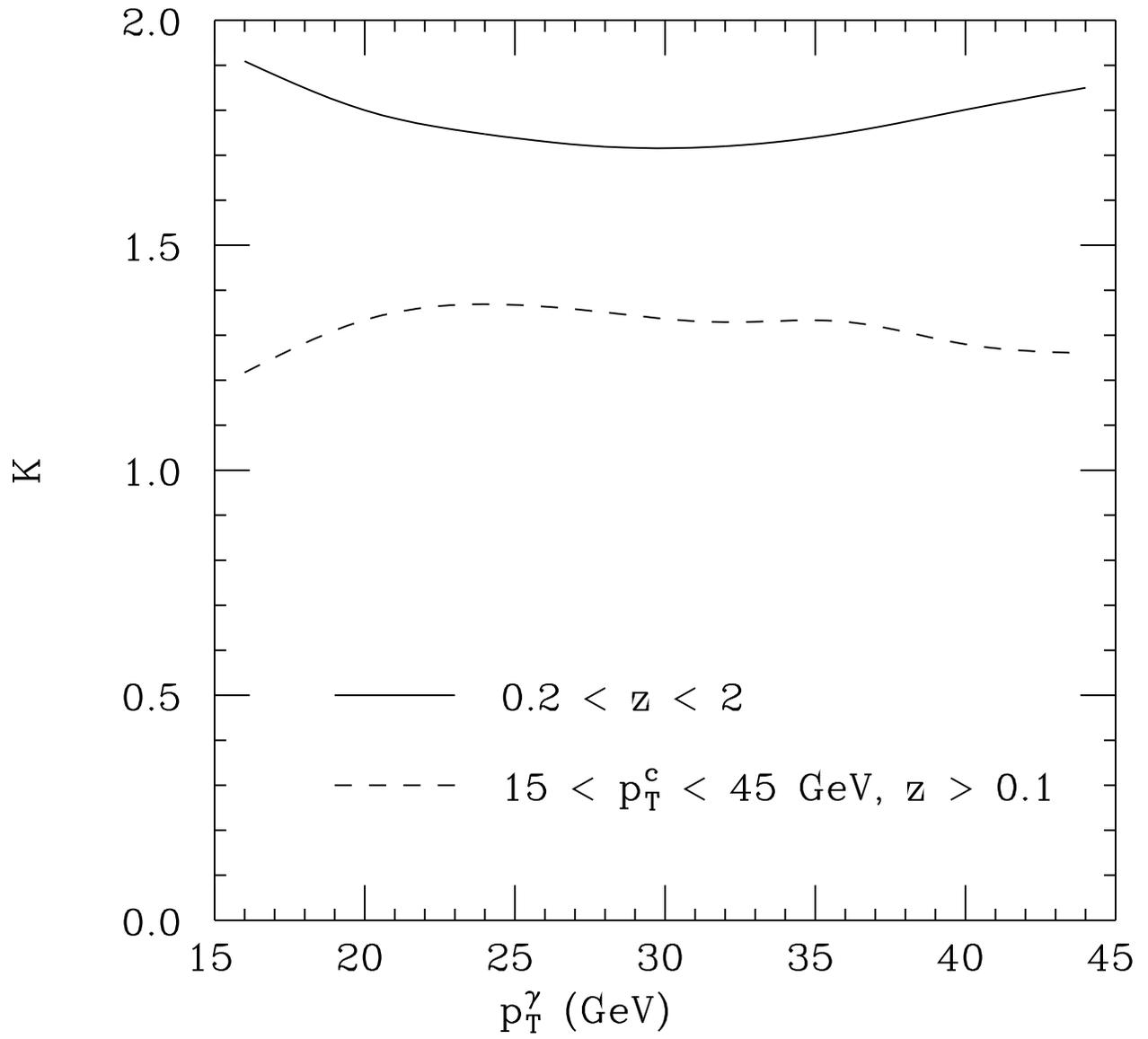
\end{document}